\documentclass[graybox, secnum]{svmult}

\usepackage{mathptmx}       
\usepackage{helvet}         
\usepackage{courier}        
\usepackage{type1cm}        
\usepackage{makeidx}         
\usepackage{graphicx}        
\usepackage{multicol}        
\usepackage[bottom]{footmisc}
\usepackage{hyperref}        
\usepackage{soul}            
\usepackage{amssymb}         
\usepackage{amsmath}         
\hypersetup{colorlinks=true,urlcolor=blue}
\usepackage[square,numbers]{natbib}
\makeindex             

\def\simgt{\lower 2pt \hbox{$\, \buildrel {\scriptstyle >}\over {\scriptstyle \sim}\,$}}
\def\simlt{\lower 2pt \hbox{$\, \buildrel {\scriptstyle <}\over {\scriptstyle \sim}\,$}}

\def\aox{$\alpha_{\rm ox}$}

\def\art{{ART-XC}}
\def\asca{{\it ASCA\/}}
\def\athena{{\it Athena\/}}
\def\axis{{\it AXIS\/}}
\def\chandra{{\it Chandra\/}}
\def\einstein{{\it Einstein\/}}
\def\ep{{\it Einstein Probe\/}}
\def\erosita{{eROSITA}}
\def\euclid{{\it {\it Euclid}\/}}
\def\hexp{{\it {\it HEX-P}\/}}
\def\hst{{\it {\it HST}\/}}
\def\integral{{\it {\it INTEGRAL}\/}}
\def\jwst{{\it {\it JWST}\/}}
\def\lynx{{\it {\it Lynx}\/}}
\def\nustar{{\it NuSTAR\/}}
\def\roman{{\it Roman\/}}
\def\rosat{{\it ROSAT\/}}
\def\sax{{\it BeppoSAX\/}}

\def\srg{{\it Spektrum-Roentgen-Gamma\/}}
\def\srgs{{\it SRG\/}}
\def\starx{{\it STAR-X\/}}
\def\swift{{\it Swift\/}}
\def\xuntian{{\it Xuntian\/}}
\def\xmm{{\it XMM-Newton\/}}

\def\xray{{\hbox{X-ray}}}
\def\xrays{{\hbox{X-rays}}}

\newcommand{\ang}{\textup{\AA}}
\newcommand{\luvr}{L_{\nu, 2500 \ang}}
\newcommand{\lxr}{L_{\rm \nu, 2 keV}}
\newcommand{\sigone}{\rm \Sigma_1}

\begin{document}
\title*{Surveys of the Cosmic X-ray Background}
\author{W.N. Brandt\thanks{corresponding author} and G. Yang}
\institute{Department of Astronomy \& Astrophysics, 525 Davey Lab, The Pennsylvania State University, University Park, PA 16802, USA;
Institute for Gravitation and the Cosmos, The Pennsylvania State University, University Park, PA 16802, USA; 
Department of Physics, 104 Davey Laboratory, The Pennsylvania State University, University Park, PA 16802, USA; \email{wnbrandt@gmail.com}
\and
Department of Physics and Astronomy, Texas A\&M University, College Station, TX 77843-4242, USA; 
George P.\ and Cynthia Woods Mitchell Institute for Fundamental Physics and Astronomy, Texas A\&M University, College Station, TX 77843-4242, USA; 
\email{gyang206265@gmail.com}}
%
%
\maketitle

\abstract{We provide a highly concise overview of what \xray\ surveys and their
multiwavelength follow-up have revealed about the nature of the cosmic \xray\
background (CXRB) and its constituent sources.
We first describe early global studies of the CXRB, the development of
imaging CXRB surveys, and the resolved CXRB fraction.
Second, we detail the sources detected in CXRB surveys describing their
identification, classification, and basic nature.
Third, since active galactic nuclei (AGNs) are the main contributors to the
CXRB, we discuss some key insights about their demographics, physics, and ecology
that have come from CXRB surveys.
Finally, we highlight future prospects for the field.}

\vspace{0.5in}

\noindent
{\bf Keywords}
Surveys;
Cosmic X-ray background;
Cosmology: observations;
Galaxies: active;
Galaxies;
Galaxies: evolution;
Galaxies: clusters;
Galaxies: groups; 
\xray\ astronomy


\section{\textit{Introduction}}

\subsection{The Cosmic X-ray Background and Early Global Studies}

The cosmic \xray\ background (CXRB) was the first cosmic background discovered (Giacconi et~al.\ 1962).
Broad-band spectral measurements showed the CXRB peaked in intensity at \hbox{20--40~keV} (see Fig.~1),
and that its \hbox{3--50~keV} spectrum could be acceptably fit with a thermal bremsstrahlung model associated
with a hot, optically thin plasma with a temperature of $kT\approx 40$~keV (e.g., Marshall et~al.\ 1980). 
Furthermore, early all-sky surveys showed the CXRB was highly isotropic, indicating that it was
primarily extragalactic in nature (e.g., Schwartz 1980). These results led some to argue that the
CXRB largely originates in a hot intergalactic medium. However, such early interpretations were
definitively ruled out by tight constraints upon the Compton-scattering distortion of the
spectrum of the cosmic microwave background (e.g., Mather et~al.\ 1990). 

Contemporaneously with these early global studies of the CXRB, increasingly sensitive investigations with
\xray\ observatories were showing that many extragalactic objects were significant \xray\ emitters,
including active galactic nuclei (AGNs), galaxies, and galaxy clusters and groups (e.g., Giacconi 1981). 
These findings indicated that at least some of the CXRB must arise from discrete sources, motivating the
formation and growth of an industry investigating the discrete-source contributions to the CXRB.

\subsection{Imaging Surveys of the CXRB: A Very Brief Review}

The discrete-source nature of much of the low-energy CXRB was supported as imaging \xray\ survey
observations with Wolter telescopes, such as \einstein\ and \rosat, resolved
increasing fractions of the CXRB below \hbox{2--3~keV} (see Fig.~2 left). 
For example, long-exposure \rosat\ surveys in the early-to-mid 1990's reached
\hbox{0.5--2~keV} flux limits of $\approx 10^{-15}$~erg~cm$^{-2}$~s$^{-1}$ and resolved
$\approx 75$\% of the soft CXRB into discrete sources
with a sky density of \hbox{800--900~deg$^{-2}$}, the majority of which
were identified as AGNs (e.g., Hasinger et~al.\ 1998; Schmidt et~al.\ 1998).
The nature of the CXRB near its high-energy peak in intensity, however, remained much less clear.
The unobscured and mildly obscured AGNs mostly discovered in early low-energy surveys did not
appear to have the requisite spectral properties to explain the broad-band CXRB spectrum,
particularly its \hbox{20--40~keV} peak. However, it was recognized that additional highly
obscured AGNs might be capable of explaining the CXRB spectrum, as their photoelectric-absorption
cutoffs would lie at higher energies. Most notably, Setti \& Woltjer (1989) put forward one of
the first ``AGN synthesis models'' for the CXRB making this point quantitatively, where reasonable
assumptions about, e.g., AGN obscuration levels, luminosities, and redshift evolution were shown
to be plausibly capable of producing the CXRB spectrum. Such synthesis models, continually refined
in light of new data, have become a mainstay of CXRB surveys research, aiding with the
interpretation of observational results and making testable predictions for new observations. 


\begin{figure}[t!]
\hspace{0.5in}
\includegraphics[height=2.4in,width=3.6in,angle=0]{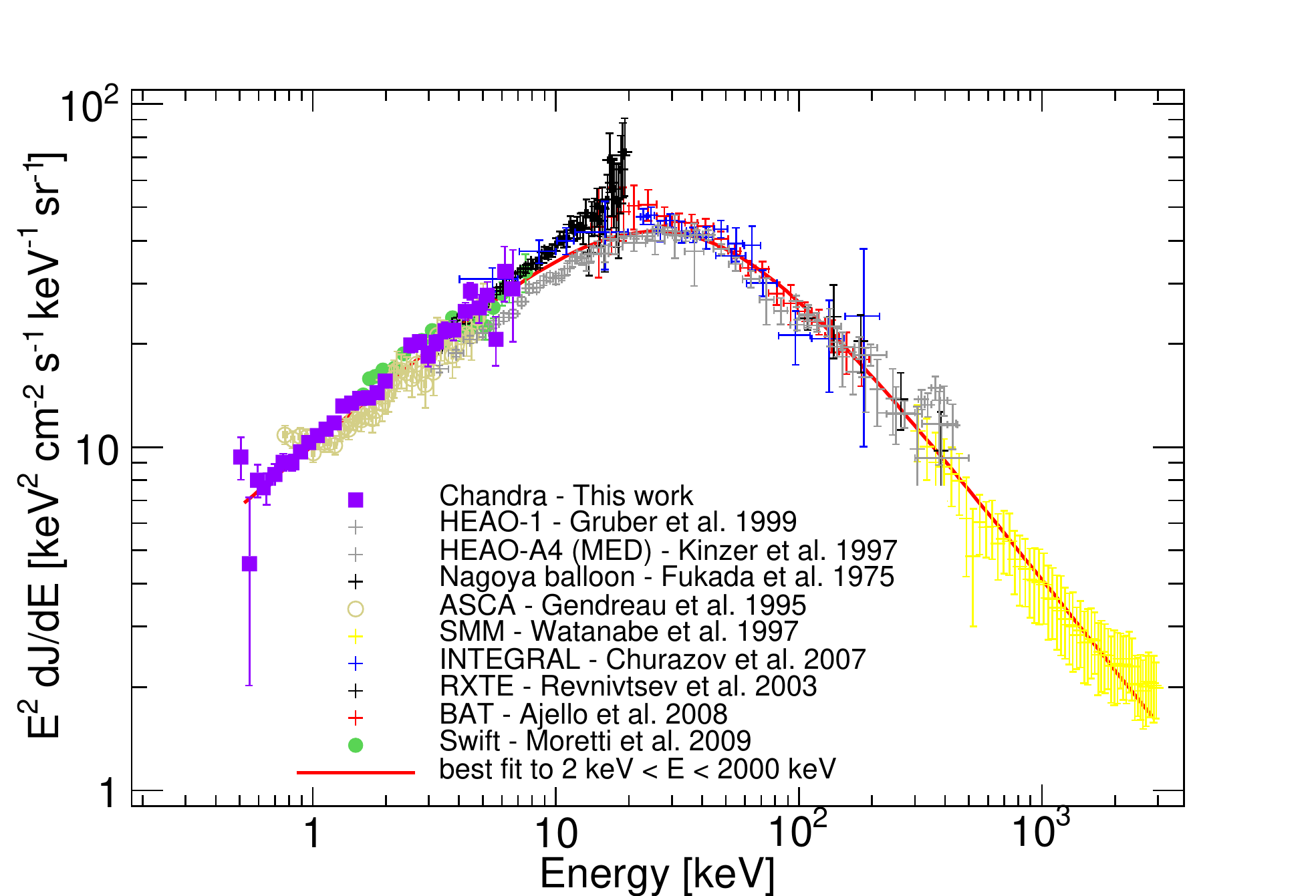}
\caption[]{Broad-band spectrum of the CXRB measured by several missions and
experiments as labeled. Note the peak at \hbox{20--40~keV}.
From Cappelluti et~al.\ (2017).}
\end{figure}


The sensitive CXRB-surveys industry at energies above \hbox{2--3~keV} took longer to
develop, largely owing to the technological challenges of focusing and detecting higher
energy \xrays. The \asca\ mission performed multiple CXRB surveys reaching limiting
\hbox{2--10~keV} fluxes of $\approx 5\times 10^{-14}$~erg~cm$^{-2}$~s$^{-1}$ at best,
detecting up to $\approx 100$ sources deg$^{-2}$ (e.g., Ueda et~al.\ 1998). The most
sensitive \asca\ surveys were source-confused owing to the limited available
angular resolution, making robust source identifications often difficult. The \sax\
mission advanced CXRB surveys at \hbox{5--10~keV}, owing to its improved high-energy
imaging capabilities, resolving \hbox{20--30\%} of the \hbox{5--10~keV} CXRB
(e.g., Comastri et~al.\ 2001). 


\begin{figure}[t!]
%
%
%
\hspace{0.5in}
\vbox{\includegraphics[height=3.1in,width=3.6in,angle=0]{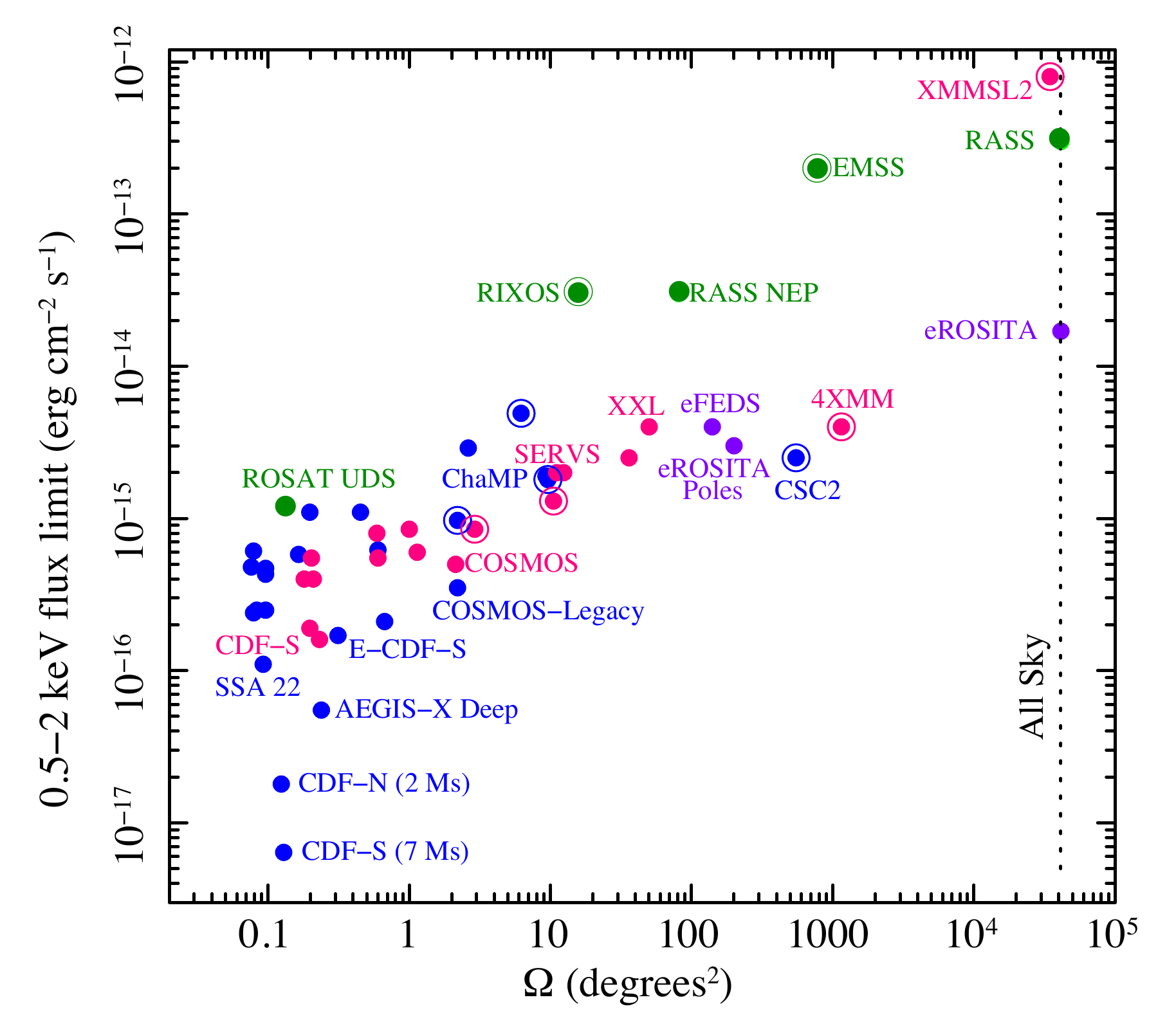}\\
\includegraphics[height=3.1in,width=3.6in,angle=0]{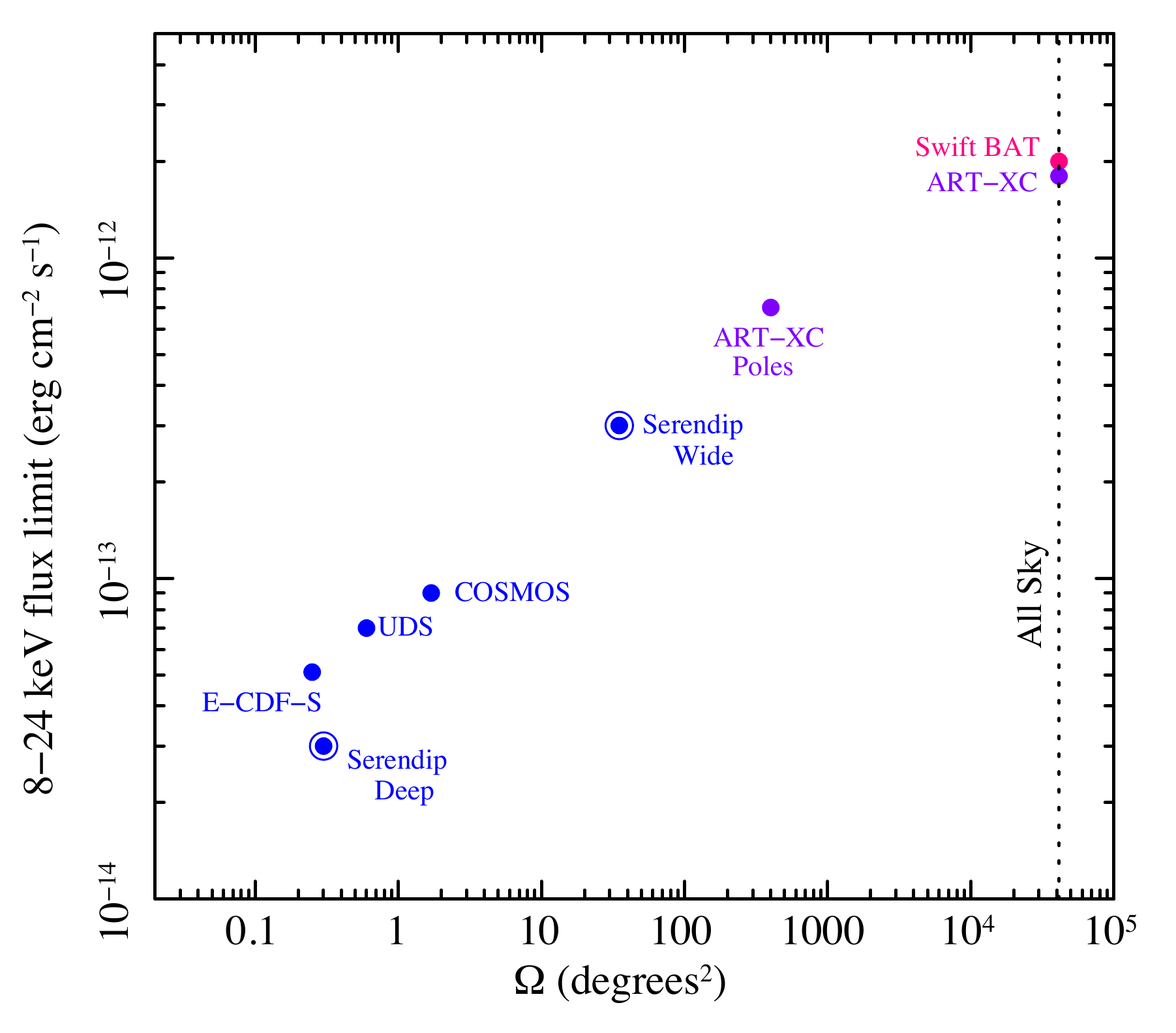}}
\caption[]{
{\it Left:\/} 
Flux limit vs.\ solid angle of sky coverage for \hbox{0.5--2~keV} surveys from
\einstein\ and \rosat\ (green),
\chandra\ (blue),
\xmm\ (red), and
\srgs\ \erosita\ (purple). 
Encircled points represent serendipitous surveys.
Some of the surveys are labeled by name in regions where symbol crowding allows. 
Updated from Brandt \& Alexander (2015), which provides additional
relevant details. 
{\it Right:\/} 
Flux limit vs.\ solid angle of sky coverage for \hbox{8--24~keV} surveys from
\swift\ BAT (red),
\nustar\ (blue), and 
\srgs\ \art\ (purple).}
\end{figure}


The superb \xray\ telescopes and detectors on \chandra\ and \xmm, both launched
in 1999, revolutionized CXRB surveys over
the \hbox{$\approx 0.5$--10~keV} band, allowing multiple powerful
surveys with up to \hbox{$\approx 150$--600} times (depending upon energy band)
the sensitivity of previous \xray\ missions (see Fig.~2 left; e.g., Brandt \& Alexander 2015). 
Large samples of up to $\approx 12,000$ sources per survey
allowed robust statistical characterization of CXRB source populations (see Fig.~3 left),
and, owing to small \hbox{0.3--3$^{\prime\prime}$} \xray\ source positional uncertainties,
the majority of the sources could be reliably matched to multiwavelength counterparts
for further characterization. The surveying capabilities of \chandra\ and \xmm\
are, moreover, highly complementary; e.g., \chandra\ excels at the highest sensitivity
surveys (see Fig.~2 left) owing to
its sub-arcsecond imaging which provides a small source-detection
cell and largely avoids source confusion, while the larger photon-collecting
area of \xmm\ provides critical spectroscopic information for sources above
its confusion limit. While \chandra\ and \xmm\ still did not directly cover
the \hbox{20--40~keV} CXRB peak, they did reveal many highly obscured AGNs
with strong signatures of nuclear Compton ``reflection''. This added  
support to the idea that such objects are largely responsible for the overall
CXRB spectrum, as formalized with improved AGN synthesis models
(e.g., Gilli et~al.\ 2007; Ananna et~al.\ 2019). 

Over the past two decades, CXRB surveys at higher energies of \hbox{10--200~keV}
also advanced dramatically (see Fig.~2 right). First, \integral\ and \swift\ BAT (launched
in 2002 and 2004, respectively) provided new coded-aperture all-sky hard \xray\ surveys
that generated sizable AGN samples, mostly at $z\simlt 0.5$, with much reduced obscuration
bias (e.g., Mereminskiy et~al.\ 2016; Oh et~al.\ 2018). Then, in 2012,
\nustar\ provided the first high-energy focusing \xray\ telescopes
in orbit, and it became possible to conduct multiple CXRB surveys reaching up
to observed \hbox{$\approx 16$--24~keV} that explore the $\simgt 10$~keV sky with
about a one-hundred-fold improvement in sensitivity
(e.g., Harrison et~al.\ 2016; Lansbury et~al.\ 2017). 
These reached ever closer to the \hbox{20--40~keV}
CXRB peak, still further confirming the presence of obscured
AGNs capable of explaining the CXRB spectrum. The AGNs detected in \nustar\
CXRB surveys mostly lie at \hbox{$z\approx 0.2$--3}, and the combination of
\integral, \swift\ BAT, and \nustar\ allow AGNs to be surveyed at high
rest-frame energies over most of cosmic time (see Fig.~3 right). 


\begin{figure}[t!]
%
\includegraphics[height=2.3in,width=2.3in,angle=0]{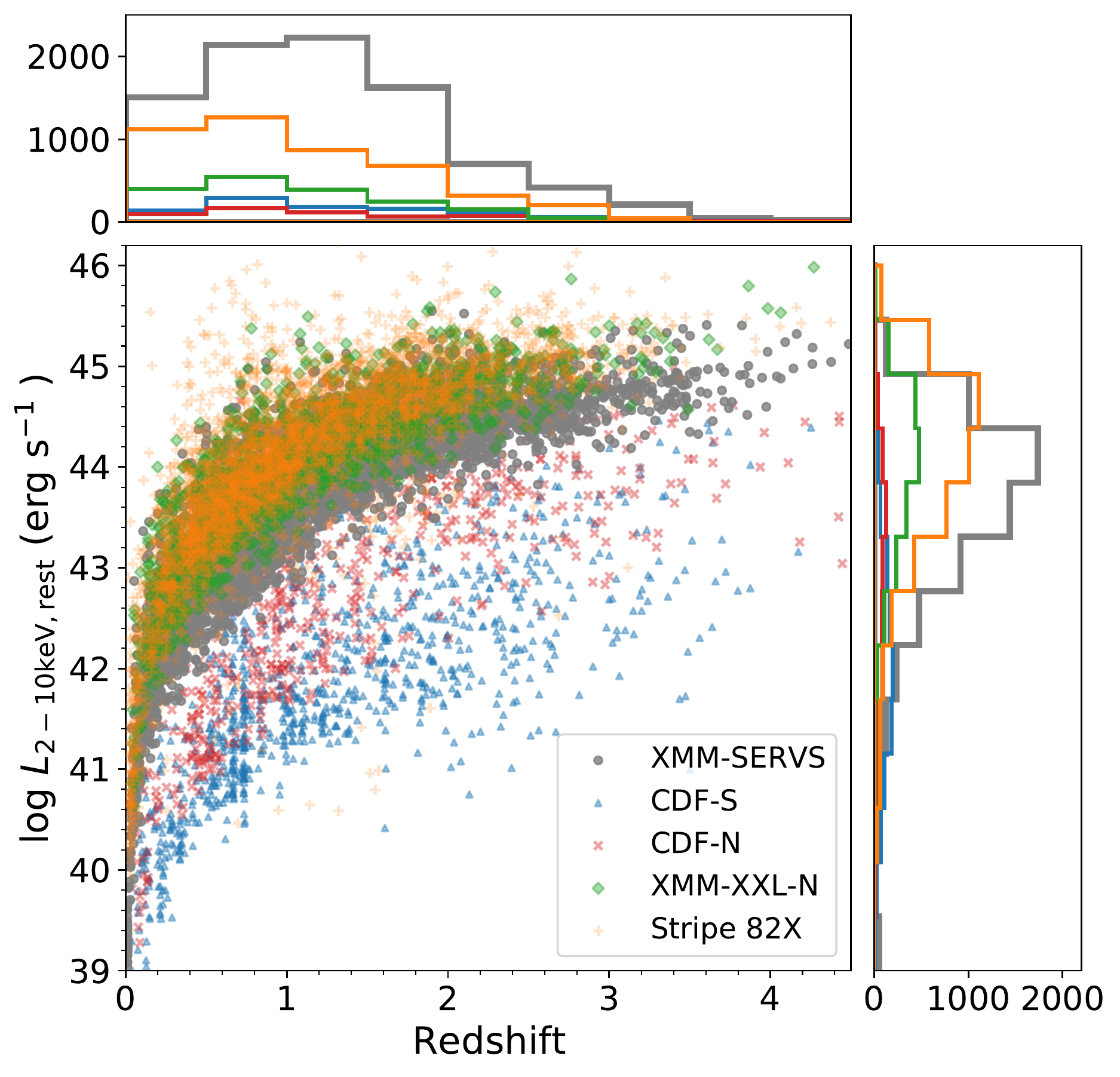}
\includegraphics[height=1.9in,width=2.3in,angle=0]{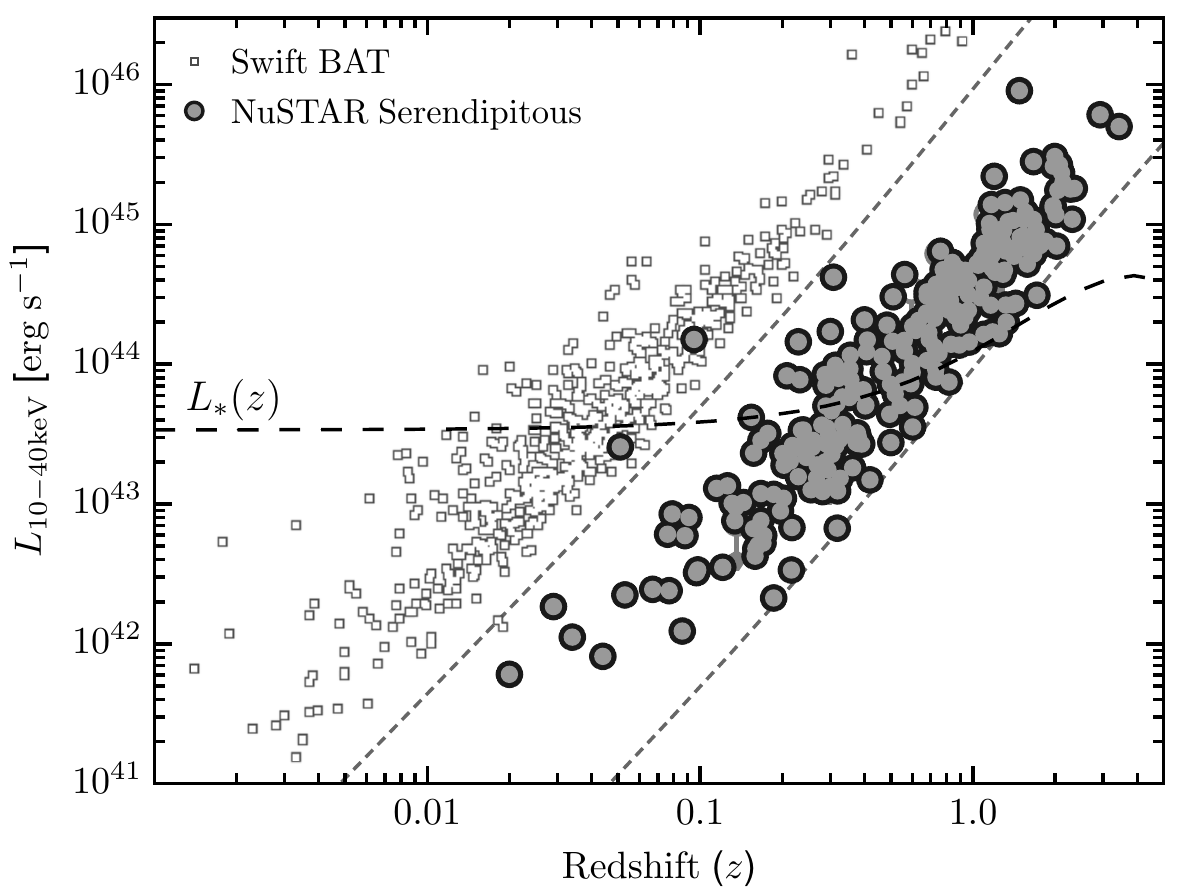}
\caption[]{
{\it Left:\/} 
Distribution of \hbox{2--10~keV} luminosity vs.\ redshift for \xray\ sources, mostly AGNs, with
spectroscopic or high-quality photometric redshifts from several \chandra\ and \xmm\ surveys
(as labeled). The top and right histograms show redshift and luminosity distributions, respectively. 
Note the large sample sizes and wide coverage of luminosity-redshift space. 
From Ni et~al.\ (2021a). 
{\it Right:\/} 
Distribution of \hbox{10--40~keV} luminosity vs.\ redshift for \xray\ sources, mostly AGNs, with
spectroscopic or high-quality photometric redshifts from the \nustar\ Serendipitous Survey and
the \swift\ BAT survey (as labeled). The two gray short-dashed lines indicate an observed \xray\ flux range
covering two orders of magnitude, from $2\times 10^{-12}$ to $2\times 10^{-14}$~erg~cm$^{-2}$~s$^{-1}$.  
The black long-dashed line indicates the evolution of the ``knee luminosity'' of the \xray\ luminosity
function $(L_\ast)$. Note the wide coverage of luminosity-redshift space. 
From Lansbury et~al.\ (2017).} 
\end{figure}


Most recently, in Fall 2019 \erosita\ on \srg\ (\srgs) started its 4-year program of
surveying the entire \hbox{0.2--5~keV} sky (e.g., Predehl et~al.\ 2021). When completed,
\erosita\ will reach typical flux limits below 2~keV (a \hbox{0.5--2~keV} flux limit of
$\approx 1.5\times 10^{-14}$~erg~cm$^{-2}$~s$^{-1}$)
about 25 times deeper than those of \rosat,
which provided the previously most-sensitive soft \xray\ all-sky survey
(see Fig.~2 left). Millions of 
extragalactic sources are expected in the final \erosita\ survey.
After these are well characterized, an enormous undertaking, they
will allow studies of the bright source populations of the CXRB
with overwhelming statistics. 
\srgs\ also carries \art, which is simultaneously conducting a sensitive
\hbox{4--30~keV} all-sky survey that should ultimately surpass other comparable 
surveys in this band considering angular resolution, sensitivity, and
uniformity together (see Fig.~2 right); the \art\ survey should detect
thousands of sources (e.g., Pavlinsky et~al.\ 2021). 

As shown in Fig.~2, the surveys conducted by the above-described missions have
effectively covered much of the \hbox{0.5--24~keV} sensitivity vs.\ solid-angle
``discovery space'' in a ``wedding-cake'' pattern (i.e., ranging from
shallow, wide-field surveys to deep, pencil-beam surveys). 

\subsection{The Currently Resolved CXRB Fraction}

For almost 60 years, a key driver of \xray\ astronomy has been to resolve the CXRB. 
While this has not yet been fully accomplished, particularly at the \hbox{20--40~keV}
CXRB peak, enormous progress has been made, and it seems clear that the CXRB
indeed primarily arises from discrete point sources, largely obscured AGNs.


\begin{figure}[t!]
%
\includegraphics[height=2.6in,width=2.4in,angle=0]{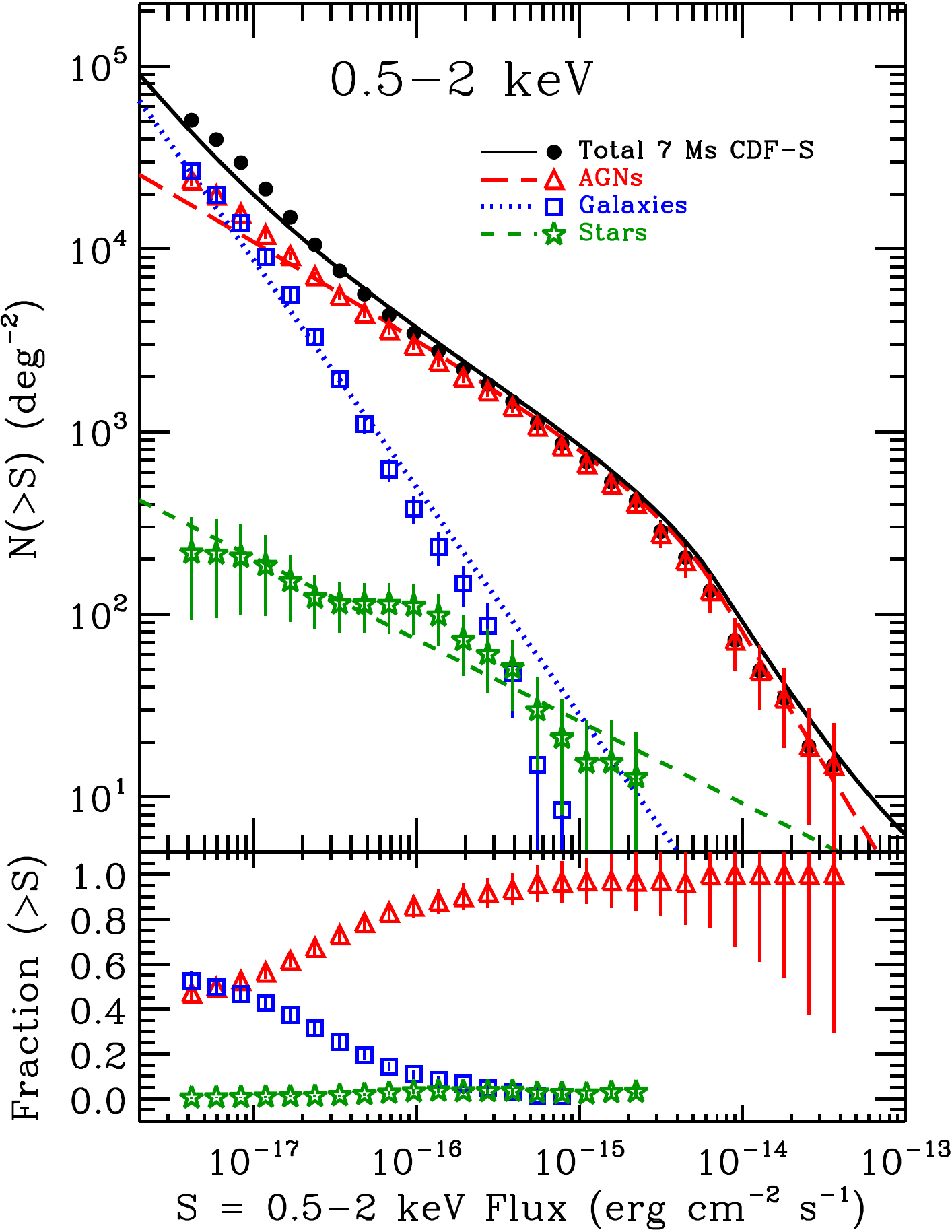}
\includegraphics[height=2.2in,width=2.4in,angle=0]{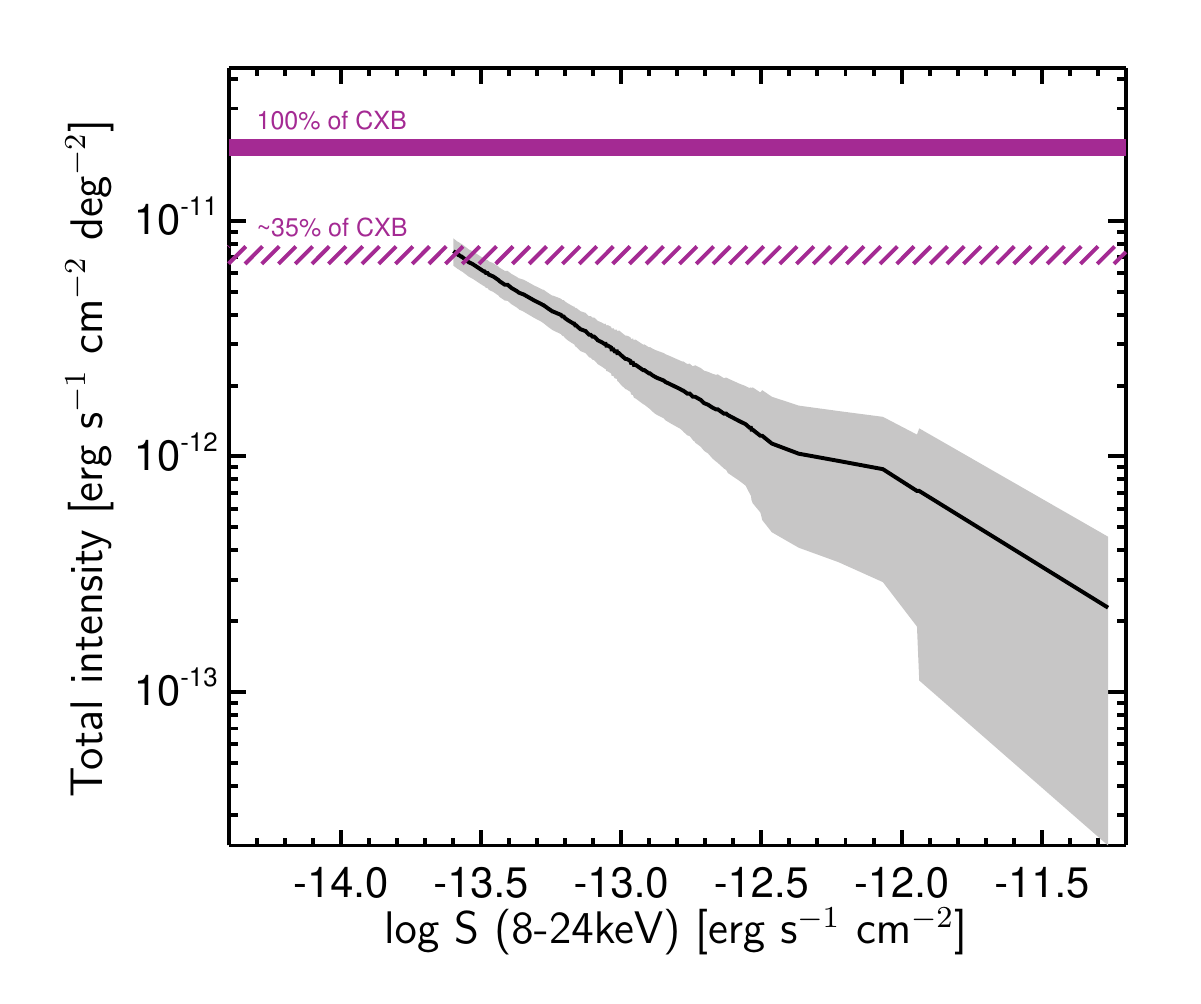}
\caption[]{
{\it Left:\/} 
Cumulative \hbox{0.5--2~keV} number counts for the 7~Ms exposure of the
\chandra\ Deep Field-South broken down by source type as labeled. The bottom
panel shows the fractional contributions to the total number counts coming
from AGNs, galaxies, and stars. Note the large sky densities of \xray\ sources,
AGNs, and galaxies detected, as well as the rapid rise of the galaxy number
counts at the faintest fluxes. 
From Luo et~al.\ (2017). 
{\it Right:\/} 
Flux contribution from resolved \nustar\ sources to the intensity of the 8--24~keV
CXRB, plotted as a function of \hbox{8--24~keV} flux. At faint fluxes, \nustar\ is
resolving $\approx 35$\% of the CXRB.
From Harrison et~al.\ (2016).}
\end{figure}


The deepest \chandra\ surveys, aided by a ``bright-end correction'' from wider
field surveys, resolve $\approx  80.9\pm 4.4$\% and $\approx 92.7\pm 13.3$\% of the CXRB
intensity in the \hbox{0.5--2~keV} and \hbox{2--8~keV} bands, respectively
(e.g., Luo et~al.\ 2017). The \chandra\ source number counts reach
$\approx 50,500$~deg$^{-2}$ with securely identified AGNs making up
$\approx 23,900$~deg$^{-2}$ and the strong majority of the resolved flux (see Fig.~4 left). 
Notably, at the faintest \hbox{0.5--2~keV} fluxes galaxies slightly surpass the sky
density of AGNs, and the galaxy number counts continue rising quickly to
the faintest flux levels probed. While some of these galaxies likely contain
unrecognized AGNs, it seems clear that galaxies will be the numerically
dominant \xray\ source population at still fainter fluxes. They are
expected to contribute $\approx 5$\% to the still-unresolved portion of
the \hbox{0.5--2~keV} CXRB; the rest likely comes from Galactic contributions,
clusters and groups, and perhaps very high-redshift ($z\simgt 6$)
black holes. 

At the important higher energies, \nustar\ surveys directly resolve $\approx 35$\%
of the \hbox{8--24~keV} CXRB emission (see Fig.~4 right; e.g., Harrison et~al.\ 2016),
although most of the relevant \nustar\ sources are not detected above $\approx 16$~keV.
The high-energy \nustar\ number counts reach $\approx 120$~deg$^{-2}$ and are
strongly dominated by AGNs.
At still higher energies (e.g., 20--60~keV), only a few percent of the CXRB
has been directly resolved (e.g., Krivonos et~al.\ 2021).


\section{\textit{Sources Detected in CXRB Surveys}}

\subsection{CXRB Source Counterparts, Redshifts, and Classifications}

Once an \xray\ source has been detected in a CXRB survey, its basic nature must be
determined before it can be used for most scientific studies. In addition to the
\xray\ data themselves, such source classification
typically relies upon both multiwavelength photometric imaging data
covering as much of the source's spectral energy distribution (SED) as possible
as well as optical/infrared spectroscopic data. Such source-classification work
is a fundamental, large, challenging, and sometimes underappreciated component
of the CXRB-surveys industry.

{\it Counterpart matching:\/}
To begin, the \xray\ source must be matched correctly to its true multiwavelength
counterpart. Such matching is now generally done statistically with likelihood-ratio
or Bayesian techniques that balance matching completeness vs.\ reliability
(e.g., Sutherland \& Saunders 1992; Salvato et~al.\ 2018), and
the availability of high-quality multiwavelength imaging data considerably helps
with the matching process (e.g., in setting statistical matching priors and accommodating
counterparts with varied SEDs). The matching process depends strongly upon the
sky density of plausible counterparts, and thus matching details vary
substantially from survey-to-survey (e.g., \swift\ BAT vs. deep \chandra\ surveys). 
In the commonly encountered regime of faint (say, $i$-band magnitudes of 
\hbox{$i\approx 20$--25}) counterparts,
\xray\ sources with positional uncertainties of \hbox{1--3$^{\prime\prime}$} or better
(e.g., from \chandra\ or \xmm) allow suitable counterpart identification for most
scientific purposes, provided high-quality multiwavelength imaging is available
(e.g., Marchesi et~al.\ 2016; Luo et~al.\ 2017; Ni et~al.\ 2021a). 
In the best cases, up to $\approx 98$\% of the \xray\ sources can be matched to
counterparts. For larger positional uncertainties in the faint-counterpart regime,
the robustness of the matching quickly declines, and caution is warranted. 
In some cases, a CXRB survey with large \xray\ positional uncertainties will first
be matched to another overlapping CXRB survey with smaller positional uncertainties
before multiwavelength counterpart determination; e.g., the identification of \nustar\
sources in sky regions also having \chandra\ or \xmm\ coverage
(e.g., Lansbury et~al.\ 2017). 

{\it Redshift determination:\/}
Once a secure multiwavelength counterpart has been found, a counterpart redshift
must be determined for most scientific applications (the \xray\ data alone
generally do not provide reliable redshifts, though there are notable
exceptions). 
Ideally, this is done with a high-quality optical/infrared spectrum that 
identifies multiple spectral features, and a wide variety of \hbox{1--10~m}
telescopes, often with multi-object spectrographs, have been productively employed
for spectroscopic follow-up of CXRB survey
sources. Spectroscopic redshifts for CXRB sources can generally be obtained down
to $i$-band magnitudes of \hbox{$i=23$--24} with
long exposures on large telescopes, but at still fainter
magnitudes the spectroscopic-redshift completeness drops rapidly (e.g., see Fig.~5 left). 
In this regime, and when spectroscopic follow-up at brighter magnitudes
becomes prohibitively expensive, photometric-redshift estimates are generally
employed. After much effort, it has now become possible to derive quality
photometric redshifts (with \hbox{$\approx 1$--8\%} accuracy in $\Delta z/(1+z)$) for many of
the AGNs found in CXRB surveys that have sensitive photometric data broadly and densely
spanning the optical/infrared SED (e.g., see Fig.~5 right; e.g., Luo et~al.\ 2017; 
Salvato et~al.\ 2019; Ni et~al.\ 2021a). However, photometric redshifts for
\xray-selected AGNs generally have larger uncertainties and a higher
fraction of catastrophic mis-estimates than
those for galaxies without active nuclei. Broad-line AGNs particularly
challenge photometric-redshift analyses and have elevated uncertainties and
catastrophic mis-estimates. 
In the best-studied CXRB survey fields, the combination of spectroscopic and
photometric approaches now provide a redshift completeness above 95\%,
sufficient for reliable conclusions about CXRB source populations over
cosmic time. 


\begin{figure}[t!]
%
\includegraphics[height=2.2in,width=2.4in,angle=0]{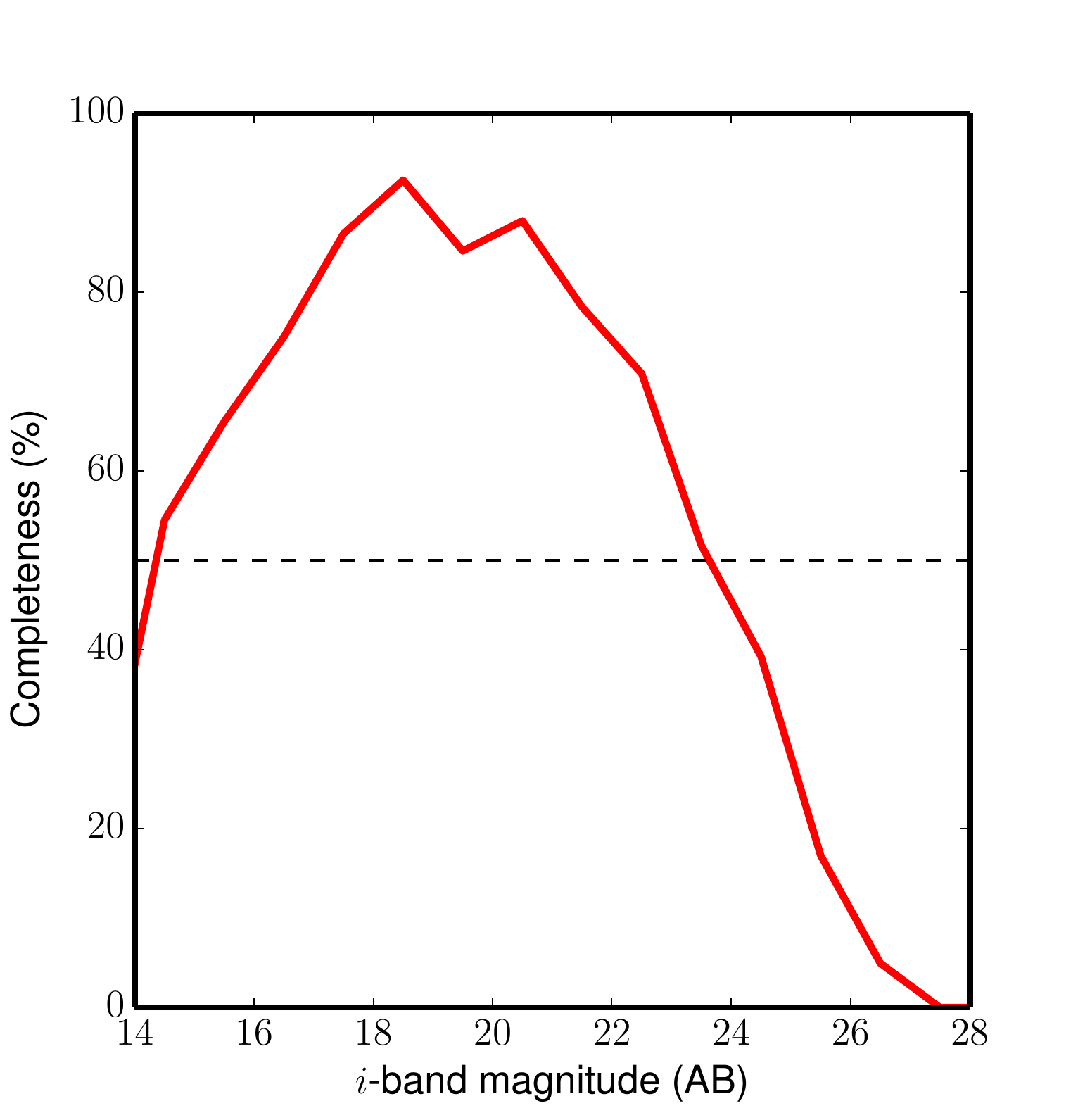}
\includegraphics[height=2.0in,width=2.2in,angle=0]{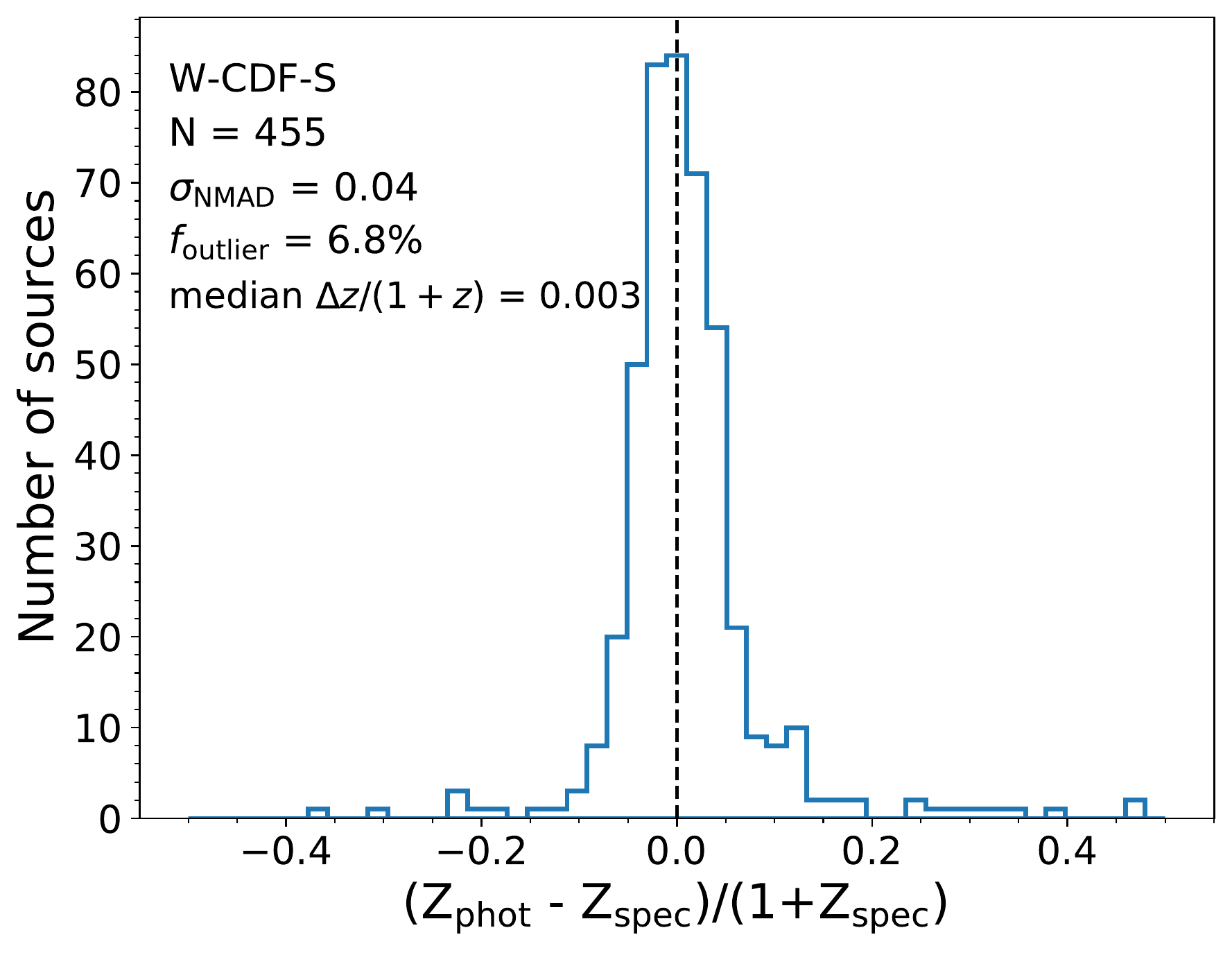}
\caption[]{
{\it Left:\/}
The spectroscopic-redshift completeness for the \chandra\ COSMOS-Legacy survey as a
function of $i$-band magnitude. Note the rapidly declining completeness toward faint 
$i$ magnitudes; this has been one persistent challenge for CXRB survey investigations
since the early 2000's (the low completeness at the brightest magnitudes arises largely 
because many of these objects are Galactic stars which were not targeted spectroscopically). 
From Marchesi et~al.\ (2016). 
{\it Right:\/}
Histogram of the fractional difference between photometric redshifts and spectroscopic
redshifts for \xray\ detected AGNs (excluding broad-line AGNs) in the 4.6~deg$^2$
region around the \chandra\ Deep Field-South (W-CDF-S). $\sigma_{\rm NMAD}$ is a robust estimator
of the standard deviation, and $f_{\rm outlier}$ is the fraction of catastrophic mis-estimates
of redshift. Overall, this is a representative example of the quality of AGN photometric
redshifts when high-quality multiwavelength data are available. 
From Ni et~al.\ (2021a).}
\end{figure}


{\it Source classification:\/}
A variety of measured properties are commonly employed to sort CXRB survey
sources having secure counterparts and redshifts into classes such as
AGNs, galaxies, clusters and groups, and transients. 
These include

\begin{enumerate}

\item
X-ray luminosity, spectral shape, variability, and morphology;

\item
X-ray-to-optical/infrared flux ratio;

\item 
Optical/infrared emission-line and continuum properties; and 

\item 
Radio morphology and core surface brightness. 

\end{enumerate}

\noindent
Further discussion of these approaches is given in, e.g., Brandt \& Alexander (2015).
Additionally, SED fitting codes can be used, e.g., to assess if significant AGN
power appears present (e.g., Pouliasis et~al.\ 2020; Yang et~al.\ 2020).
In most cases, several independent approaches are utilized
to cross-check classifications, thereby providing reliably classified samples. 

\subsection{Main Extragalactic Source Types}

Classification of CXRB survey sources reveals a wide variety of source types,
and below we will briefly describe the main extragalactic populations discovered. 

\vspace{0.1in}

{\it AGNs:\/}
AGNs are the energetically dominant source type detected in all CXRB
surveys, and they also dominate numerically in all but the \hbox{1--2} deepest
\hbox{0.5--2~keV} surveys (where galaxies appear to dominate numerically). 
Their primary \xray\ emission is thought to be largely created
via Compton up-scattering in an accretion-disk
``corona'' in the immediate vicinity of the supermassive black hole (SMBH),
but \xrays\ may also arise from the accretion disk itself (at low energies)
or jets.
CXRB surveys are extremely effective at discovering AGNs for three
key reasons:

\begin{enumerate}

\item
Luminous AGNs appear almost universally to produce substantial
\xray\ emission, typically making up \hbox{1--20\%} of the total power
emitted;

\item
The penetrating nature of high-energy \xrays\ allows detection of
the common obscured systems that are often missed at other wavelengths,
and any \xray\ biases against obscured systems decline with increasing
redshift; and

\item
X-rays provide superior contrast between AGN light and host-galaxy
starlight, thereby minimizing confusion between the two.

\end{enumerate}

\noindent
These three points together explain why deep \xray\ surveys have discovered
the highest sky density of AGNs, up to 23,900~deg$^{-2}$, compared to 
any other wavelength (see Fig.~4 left).  

The variety of CXRB surveys conducted (see Fig.~2) have discovered AGNs in significant
numbers from the local universe to $z\approx 5$, thereby allowing studies of
AGN properties over most of cosmic time (e.g., see Fig.~3). Even at high redshifts of
\hbox{$z\approx 2$--5}, deep CXRB surveys allow the discovery of many
moderate-to-low luminosity AGNs, comparable to the well-studied Seyfert
galaxies in the local universe (e.g., Vito et~al.\ 2018). 
The full range of AGN \xray\ luminosity commonly found across
CXRB surveys spans an impressive five orders of magnitude, 
\hbox{$L_{\rm X}=10^{41}$--$10^{46}$~erg~s$^{-1}$}, and optical counterpart
magnitudes span \hbox{$i_{\rm AB}\approx 14$--28}. 

As expected from the shape of the CXRB, \xray\ surveys reveal vast numbers
of obscured AGNs at all redshifts, with a generally higher fraction of
obscured AGNs being found by progressively higher energy surveys. \xray\
spectral fitting of the obscured systems finds intrinsic absorption
column densities spanning $N_{\rm H}=10^{21}$--$10^{24.5}$~cm$^{-2}$ and
frequently reveals \xray\ absorption complexity (e.g., partially covering
or ionized absorption). There are also unobscured systems
having intrinsic $N_{\rm H}$ consistent with zero. 
The AGN column density distribution generally appears log-normal in form with
a peak around $N_{\rm H}=10^{23.5}$~cm$^{-2}$ (e.g., Ueda et~al.\ 2014; 
Liu et~al.\ 2017) when averaged over wide ranges of redshift and luminosity
(note the $N_{\rm H}$ distribution depends upon both of these quantities;
see \S3.2).

Consistent with the wide ranges of AGN luminosity and \xray\ absorption level,
the optical/infrared appearances of AGNs from CXRB surveys are varied,
ranging from obvious blue, broad-line AGNs to elusive AGNs that are
host dominated and challenging to distinguish from normal galaxies
(e.g., Luo et~al.\ 2010; Hickox \& Alexander 2018). The \xray\
unobscured vs.\ obscured classifications do not always map simply onto the
classical optical type~1 vs.\ type~2 classifications, though there is
generally respectable correspondence
(e.g., Merloni et~al.\ 2014; Koss et~al.\ 2017). 

Much additional information about the AGNs from CXRB surveys is given
in \S3. 

\vspace{0.1in}


\begin{figure}[t!]
%
\includegraphics[height=2.2in,width=2.3in,angle=0]{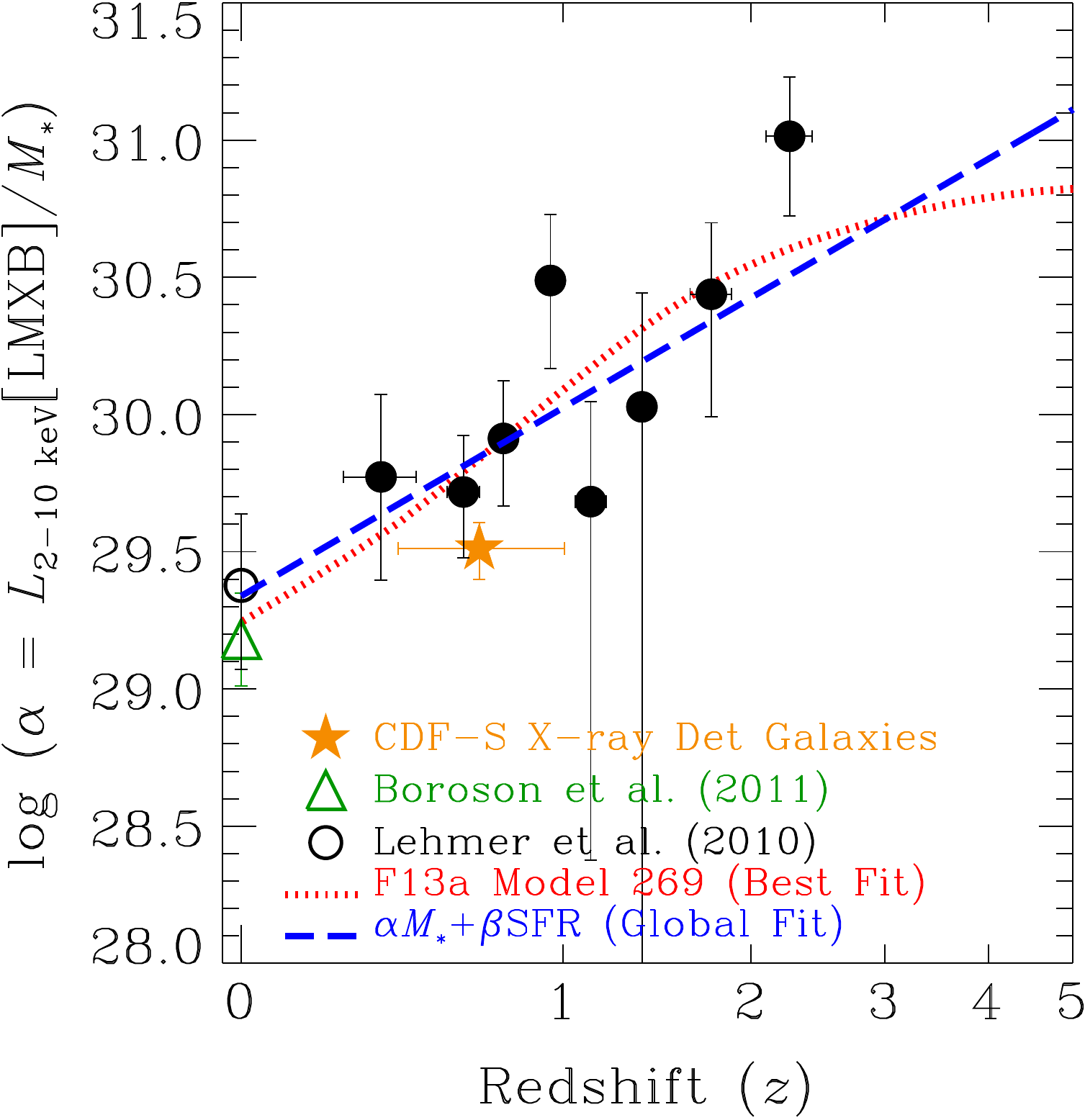}
\includegraphics[height=2.2in,width=2.3in,angle=0]{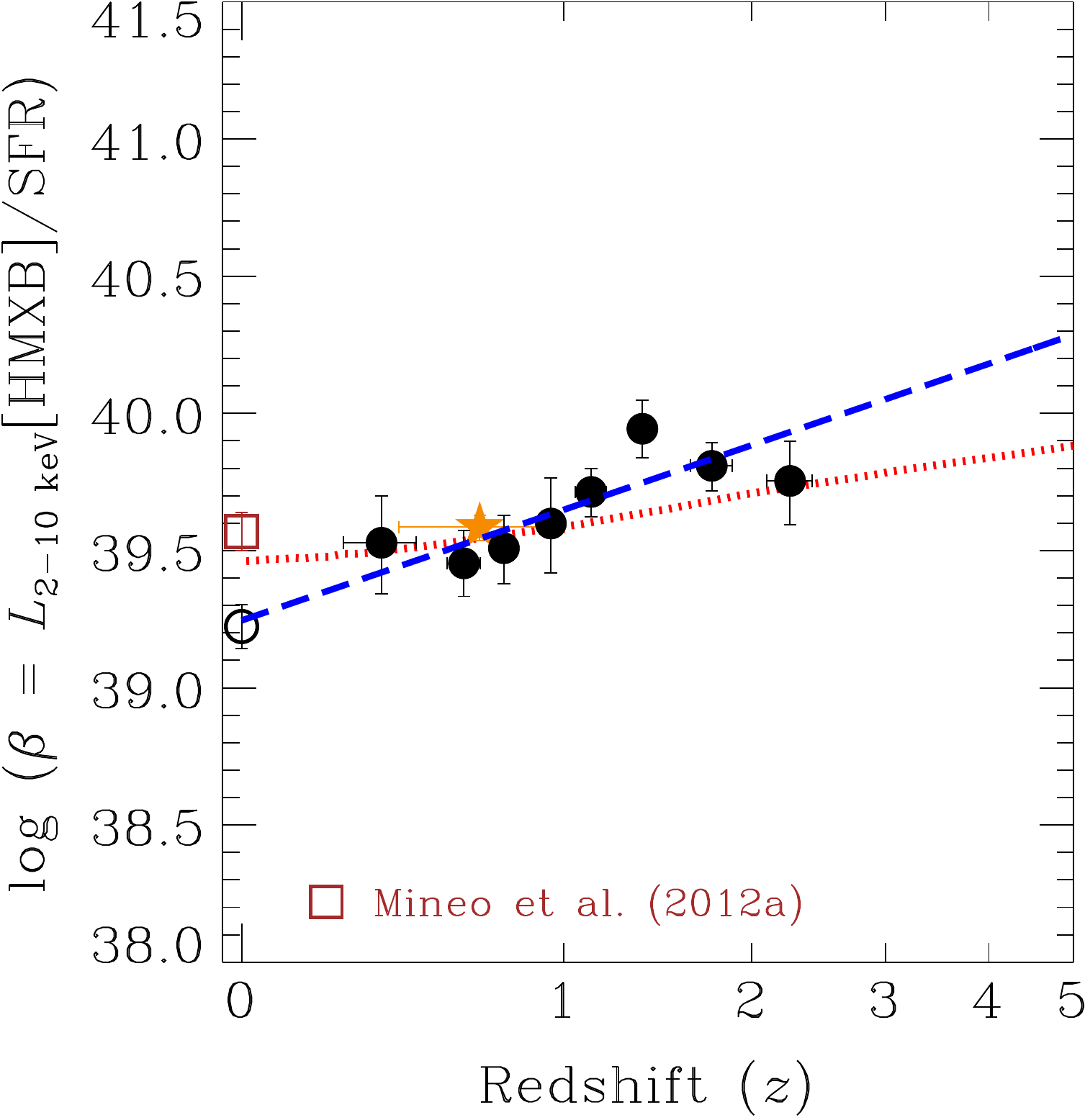}
\caption[]{Redshift dependence of
$L_{\rm 2-10\rm\ keV}({\rm LMXB})/M_\star$ and
$L_{\rm 2-10\rm\ keV}({\rm HMXB})/{\rm SFR}$  
derived from stacking analyses of \hbox{$z\approx 0$--2.5}
galaxies with sensitive \xray\ coverage (using samples as labeled). 
The dashed blue curves show the best-fit relations, and
the dotted red curves show predicted evolution from 
one population model of XBs. 
From Lehmer et~al.\ (2016).}
\end{figure}


{\it Galaxies:\/}
Galaxies are significant contributors
to the \hbox{0.5--7~keV} source number counts at the faint fluxes probed by
deep CXRB surveys. They most prominently contribute in the \hbox{0.5--2~keV}
band but also materially contribute from \hbox{2--7~keV}. Galaxies numerically
dominate over AGNs in the \xray\ number counts below \hbox{0.5--2~keV}
fluxes of $\approx 10^{-17}$~erg~cm$^{-2}$~s$^{-1}$ (see Fig.~4 left); a similar
situation is found in deep radio surveys reaching below a few mJy. 
The detected \xray\ emission primarily originates from the
accreting \xray\ binary (XB) populations inhabiting these galaxies (including
ultraluminous \xray\ sources). Below \hbox{$\approx 1$--2~keV}, hot gas emission
often contributes materially as well, arising from, e.g., supernova remnants,
starburst-driven outflows, and hot interstellar media. Unrecognized SMBH
activity (e.g., from low-luminosity or elusive highly obscured AGNs) inevitably
also contributes in some systems. 

The \xray\ detected galaxies in deep CXRB surveys typically lie at \hbox{$z\approx 0$--1.5}
(e.g., Luo et~al.\ 2017), and they thereby allow the evolution of galaxy \xray\ source
populations to be investigated over the last $\approx 70$\% of
cosmic time. These galaxies generally
have \xray\ luminosities of \hbox{$L_{\rm X}=10^{39}$--$10^{42}$~erg~s$^{-1}$} and
relatively soft \xray\ spectra with effective power-law photon indices of
\hbox{$\Gamma=1.7$--2.2}. They include both late-type and early-type
morphologies, and they span wide ranges of stellar mass ($M_\star$) 
and star-formation rate (SFR). 

Stacking-based analyses with deep CXRB surveys have allowed systematic
studies of normal-galaxy \xray\ source populations over the redshift
range \hbox{$z\approx 0$--3} (e.g., Lehmer et~al.\ 2016), and
these have shown that average galaxy $L_{\rm X}$ over cosmic time is a
function of at least $M_\star$ (primarily relating to low-mass XBs; LMXBs),
SFR (primarily relating to high-mass XBs; HMXBs), and redshift. The redshift
evolution leads to generally rising $L_{\rm X}$ at higher redshift
with scaling relations of 
$L_{\rm 2-10\rm\ keV}({\rm LMXB})/M_\star\propto (1+z)^{2-3}$ and
$L_{\rm 2-10\rm\ keV}({\rm HMXB})/{\rm SFR}\propto (1+z)$; see Fig.~6. 
The upward $L_{\rm X}$ evolution for LMXBs and HMXBs is respectively attributed
to declining galaxy stellar ages and metallicities
(e.g., Fornasini et~al.\ 2020), although further
investigations of these putative drivers are required. 


\begin{figure}[t!]
%
\includegraphics[height=2.2in,width=2.4in,angle=0]{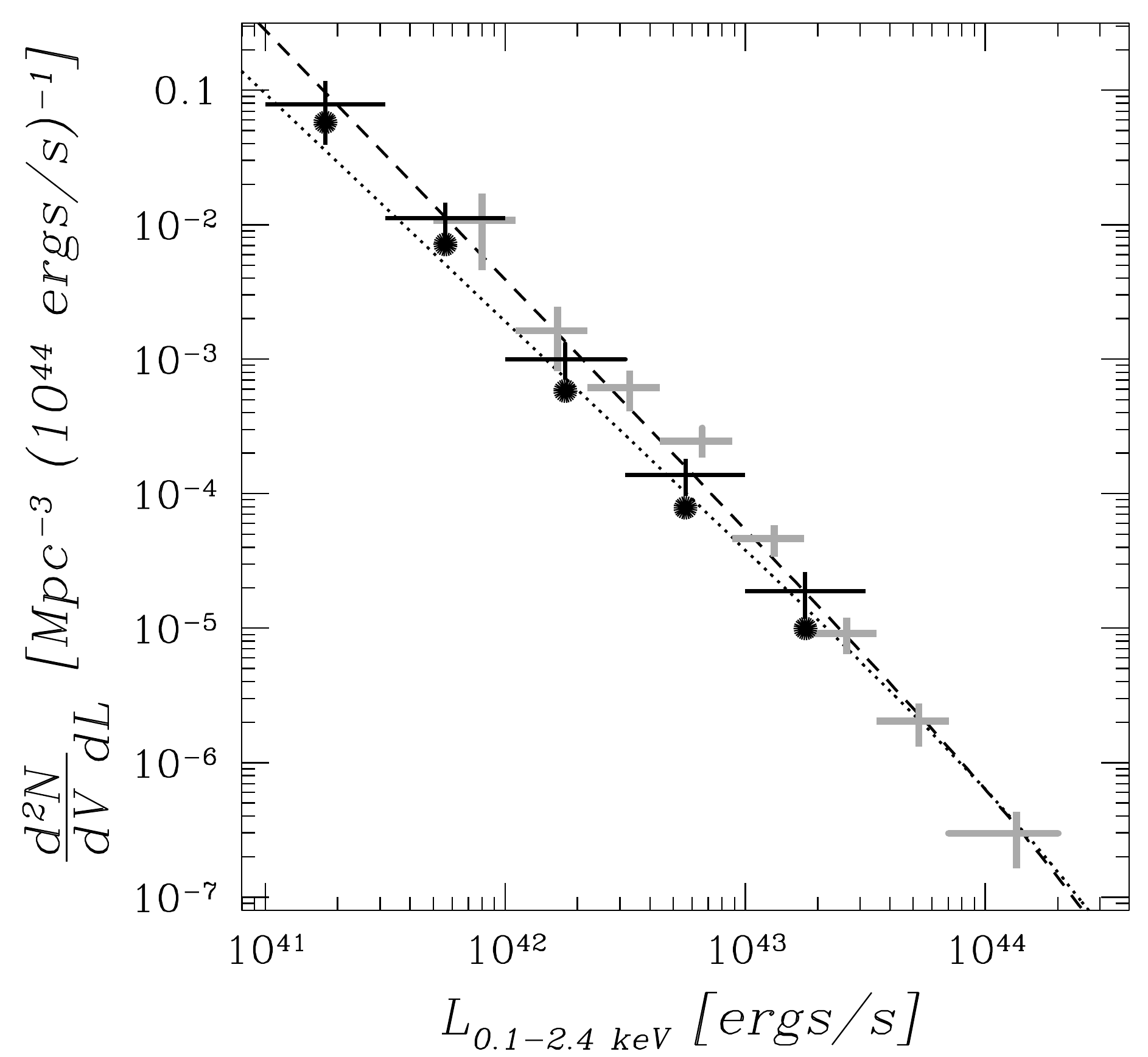}
\hspace{0.1in}
\includegraphics[height=2.4in,width=2.2in,angle=0]{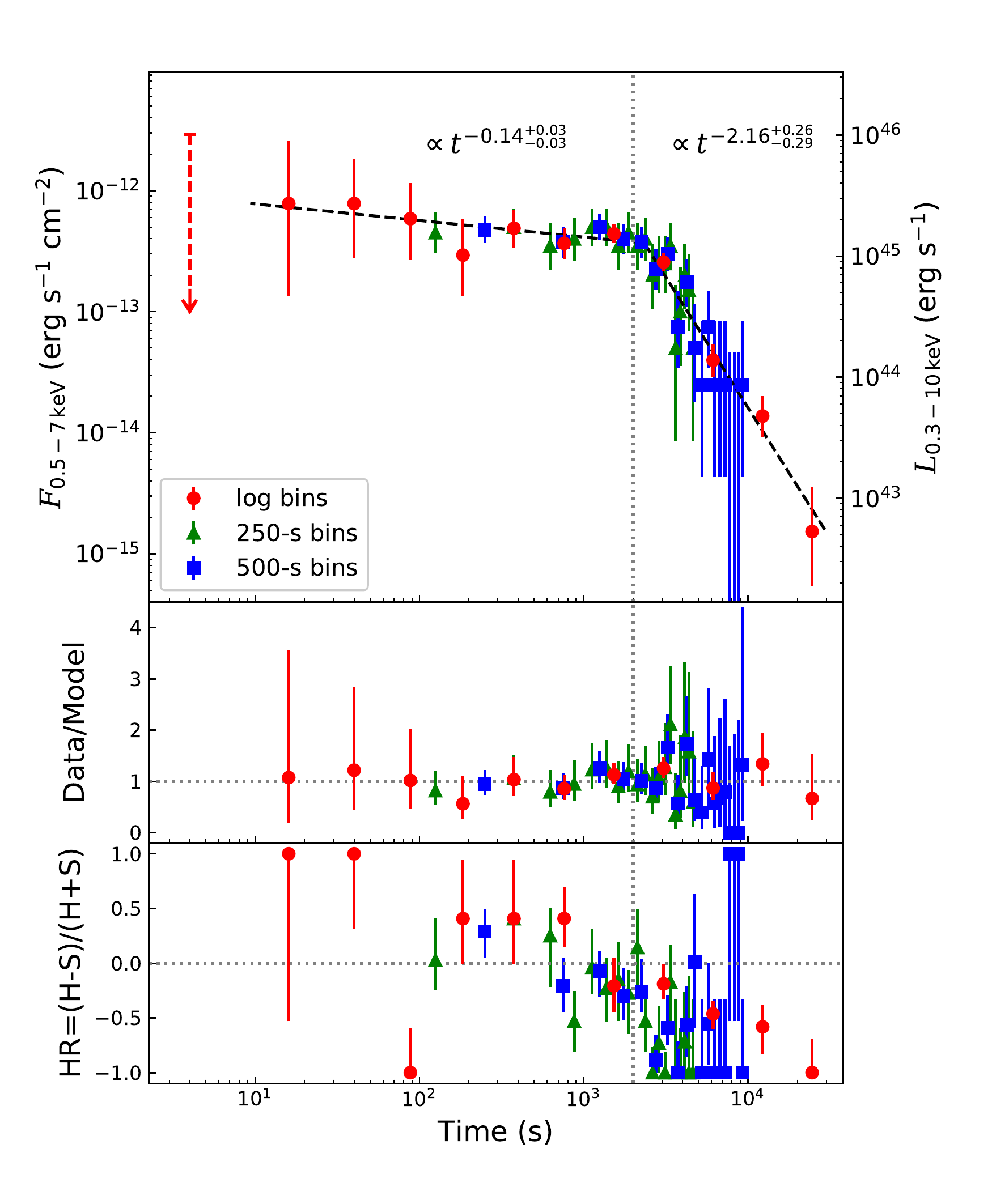}
\caption[]{
{\it Left:\/}
The \xray\ luminosity function of $z<1.2$ clusters/groups from
COSMOS (gray crosses) and 
the Extended \chandra\ Deep Field-South (black crosses and dots).
For comparison, the dashed and dotted curves show the local
\xray\ luminosity function in the Northern and Southern hemispheres,
respectively, from the \rosat\ All-Sky Survey. 
From Finoguenov et~al.\ (2015). 
{\it Right:\/}
Example \xray\ light curve of a FXRT found serendipitously in a \chandra\ CXRB
survey observation of the \chandra\ Deep Field-South. The middle panel shows
the ratio between the data and broken power-law model fit in the upper panel.
The lower panel shows the \xray\ hardness ratio. 
Based on its detailed properties, this FXRT has been proposed to be powered
by a magnetar resulting from a binary neutron star merger. 
From Xue et~al.\ (2019).}
\end{figure}


\vspace{0.1in}

{\it Galaxy clusters and groups:\/}
Galaxy clusters and groups are detected
in significant numbers (up to $\approx 400$~deg$^{-2}$)
in CXRB surveys below $\approx 10$~keV, standing out
owing to their spatial extension and relatively soft \xray\ spectra. Their
\xray\ emission arises from their hot intracluster/intragroup media (ICM/IGM)
via a combination of bremsstrahlung, line radiation, and recombination emission;
AGNs present in the cluster/group also often contribute to the total \xray\
luminosity. The ICM generally outweighs the optically visible galaxies in
clusters by a factor of \hbox{$\approx 1.5$--10}, yet it is very difficult
to observe at non-\xray\ wavelengths; in groups the IGM often has a mass
comparable to or less than that of the galaxies. 

CXRB surveys have discovered thousands of clusters/groups with 
\hbox{$L_{\rm X}\approx 10^{41}$--$10^{45}$} erg~s$^{-1}$ from \hbox{$z\approx 0$--1}
and an increasing number at \hbox{$z\approx 1$--2} and beyond (e.g., see Fig.~7 left;
e.g., B\"ohringer et~al.\ 2004; Finoguenov et~al.\ 2015; 
Adami et~al.\ 2018; Liu et~al.\ 2021a). 
These span masses of $M_{500}\approx (0.1$--20)$\times 10^{14}$~M$_\odot$ ($M_{500}$ is
the mass within $r_{500}$, the radius within which the mean density is equal to
500 times the critical density). 

Cluster/group samples from CXRB surveys have provided key insights into the
$\Lambda$CDM cosmology via measurements of, e.g., the baryonic mass
fraction and the evolution of cluster number counts (e.g., Allen \& Mantz 2020). 
The \erosita\ survey should soon deliver $\simgt 10^5$ clusters extending to
$z\simgt 1$, and these will ultimately provide precision cosmological constraints
(e.g., Predehl et~al.\ 2021). 

\vspace{0.1in}

{\it Transients:\/}
CXRB surveys have played a role in serendipitously
discovering new types of extragalactic \xray\ transients. This dates back to
at least the \rosat\ All-Sky Survey, which discovered multiple tidal disruption
event (TDE) candidates and other galactic giant \xray\ outbursts, some of which
remain poorly understood (e.g., Komossa 2015). 
\chandra\ and \xmm\ surveys have discovered other extragalactic transients,
including the revelation of a new population of fast
\xray\ transients (FXRTs) with durations of $\approx 5000$~s
and relatively faint \xray\ fluxes
(e.g., see Fig.~7 right; e.g., Yang et~al.\ 2019a; Quirola-V\'asquez et~al., in prep). 
FXRTs may be binary neutron-star merger events that form magnetars,
off-axis gamma-ray bursts, and/or 
TDEs involving the disruption of a white dwarf by an intermediate mass black hole.
The \erosita\ survey is now discovering new \xray\ selected TDEs and other
transients. 



\section{\textit{Insights on the AGN Population from CXRB Surveys}}

As described in \S2, AGNs are the energetically dominant contributors to 
the CXRB and usually dominate numerically in CXRB surveys as well. 
Therefore, in this section we will present some key insights about AGN 
demographics, physics, and ecology revealed by CXRB surveys.

\subsection{AGN Demographics}
\label{sec:demo}

\textit{Pre-Chandra/XMM-Newton era}:
The AGN luminosity function (LF) describes the AGN comoving number density as 
a function of luminosity and redshift. 
Since the discovery of the first AGNs, the LF has attracted intense investigations.  
Early studies before $\approx 1990$ were largely limited to the LF's luminous end 
(i.e., quasars, \hbox{$\approx 10$--100~deg$^{-2}$}) selected by optical surveys 
(see, e.g., Hartwick \& Schade 1990 for a review).
These studies established that the number density of quasars peaks at 
\hbox{$z\approx 2$--3}, despite the relatively large uncertainties at high redshifts 
($z\gtrsim 3$).

\rosat, thanks to its unprecedented sensitivity (at the time; see \S1.2)
from \hbox{0.1--2.4~keV} and large field of view ($2^\circ$ diameter),
delivered several key results that 
fundamentally shaped our understanding of AGN demographics. 
The \rosat\ Deep Survey detected discrete \xray\ sources reaching a surface density 
of \hbox{$\approx 800$--900~deg$^{-2}$}, most ($\gtrsim 80\%$) of which are AGNs
(e.g., Hasinger et~al.\ 1998; Schmidt et~al.\ 1998).
\rosat\ enabled the first comprehensive studies of the AGN \xray\ luminosity function 
(XLF; e.g., Miyaji et~al.\ 2000).
These favored luminosity-dependent density evolution (LDDE) models, 
such that more luminous AGNs evolve more strongly as a function of redshift.
Similar to the optical/radio results, the \rosat-based XLF rises sharply from $z=0$ 
to $z\approx 2$--3.
However, the XLF was poorly constrained at higher redshifts, and the inferred \xray\ 
AGN number density was consistent with a constant at $z\gtrsim 3$.
The XLF could account for 60--90\% (depending upon the specific LDDE model) of the 
0.5--2~keV CXRB, indicating that the observed soft CXRB largely originates from distant 
AGNs. 

\vspace{0.1in}

\textit{Redshift evolution of AGN number density:}
Thanks to their order-of-magnitude sensitivity improvements compared to previous 
missions, \chandra\ and \xmm\ have significantly pushed forward studies of AGN 
demographics.
Their deepest surveys have detected an enormously dense AGN population, reaching a
sky density of 23,900~deg$^{-2}$ (e.g., Luo et~al.\ 2017).
This density is more than 1 and 2 orders of magnitude higher than those of 
\rosat-detected AGNs and SDSS quasars, respectively (e.g., Hasinger et~al.\ 1998; 
Palanque-Delabrouille et~al.\ 2013). 

\chandra\ and \xmm\ are able to detect a fairly complete AGN sample over wide 
luminosity, redshift, and obscuration ranges (see Fig.~3 left),
allowing detailed characterization of the AGN XLF. 
The most outstanding feature of the AGN XLF is so-called ``AGN downsizing'':
the number density of higher luminosity AGNs peaks at higher redshift (e.g., 
Cowie et~al.\ 2003; Ueda et~al.\ 2014; Aird et~al.\ 2015). 
For example, AGNs with $L_X \approx 10^{45}$~erg~s$^{-1}$ and 
$L_X \approx 10^{43}$~erg~s$^{-1}$ peak at $z\approx 2.5$ and $z\approx 1$, 
respectively.
Fig.~8 (left) displays the number-density evolution for AGNs within 
different luminosity ranges, where downsizing is apparent. 
Similar downsizing behavior has also been found among optically and radio-selected 
AGNs (e.g., Croom et~al.\ 2009; Rigby et~al.\ 2011). 
The AGN-downsizing behavior suggests that more-massive SMBHs today generally form 
earlier in cosmic history. 
Interestingly, a similar phenomenon has also been found in studies of galaxy formation: 
in the local universe, more-massive galaxies tend to have older stellar populations
than less-massive ones (e.g., Thomas et~al.\ 2005);
observations of distant galaxies have directly demonstrated this galaxy-downsizing picture
(e.g., Cowie et~al.\ 1996; Juneau et~al.\ 2005).
The evolutionary similarity between AGNs and galaxies suggests some fundamental links 
between SMBH and stellar growth (see \S3.3).

AGN downsizing behavior poses a significant theoretical challenge. 
In the framework of the $\Lambda$CDM universe, dark-matter halos form 
hierarchically, i.e., low-mass halos form first and they merge into massive 
halos later in cosmic time (e.g., White \& Rees 1978; Blumenthal et~al.\ 1984).
A straightforward scenario is that SMBH and galaxy growth also follow this 
hierarchical path, because they evolve in the central regions of dark-matter halos. 
However, the observed AGN and galaxy downsizing appear to go against this scenario, 
and thus downsizing phenomena are often described as ``anti-hierarchical''.
Modern cosmological simulations can roughly yield the trend of downsizing by 
involving physical mechanisms of AGN feedback (e.g., Scannapieco et~al.\ 2005; 
Sijacki et~al.\ 2015; Volonteri et~al.\ 2016).
However, further work is needed to reproduce the observed AGN XLF across all
redshifts and luminosities (e.g., Rosas-Guevara et~al.\ 2016; Habouzit et~al.\ 2019).


\begin{figure}[t!]
\includegraphics[width=0.45\linewidth,angle=0]{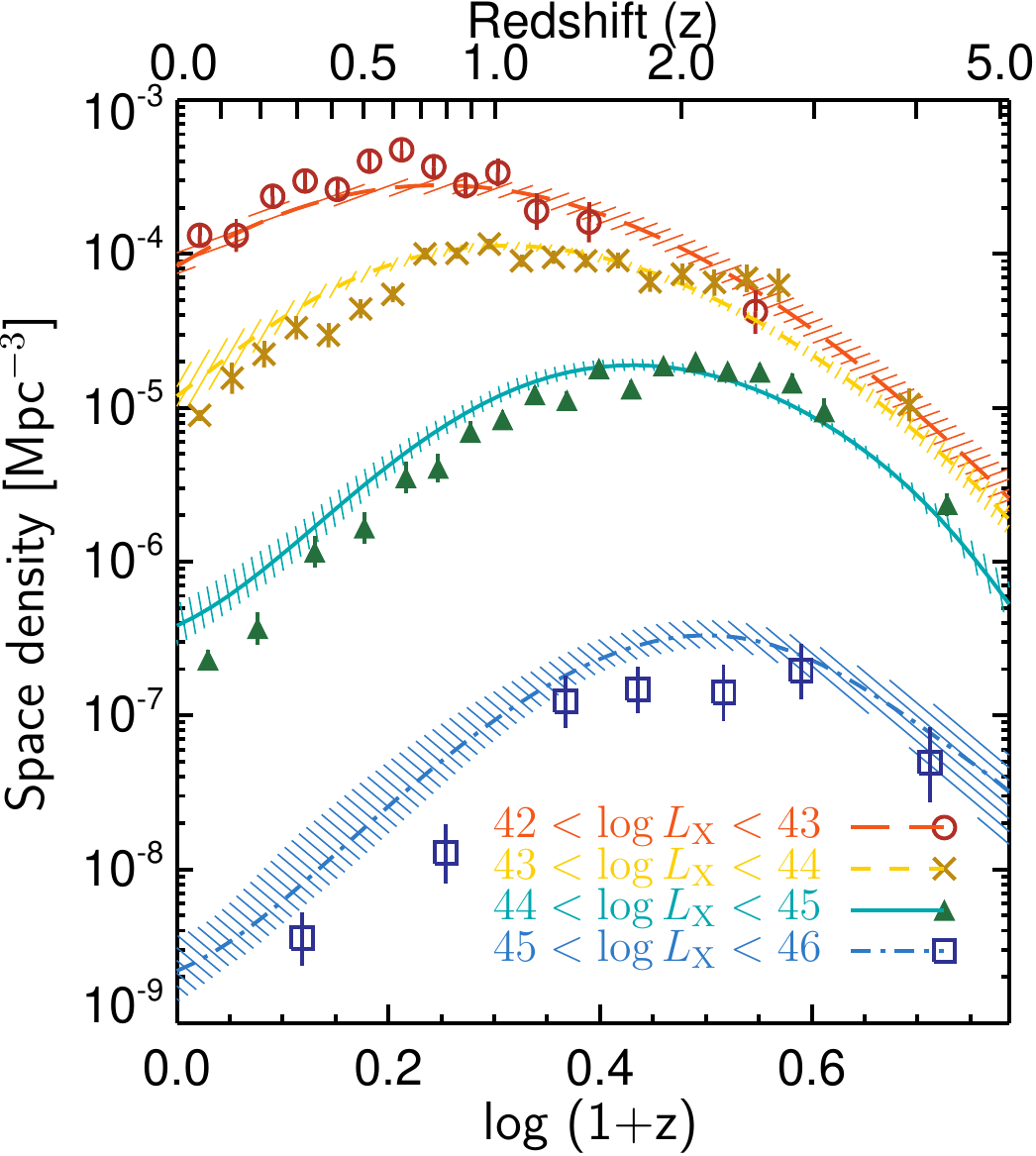}
\includegraphics[width=0.55\linewidth,angle=0]{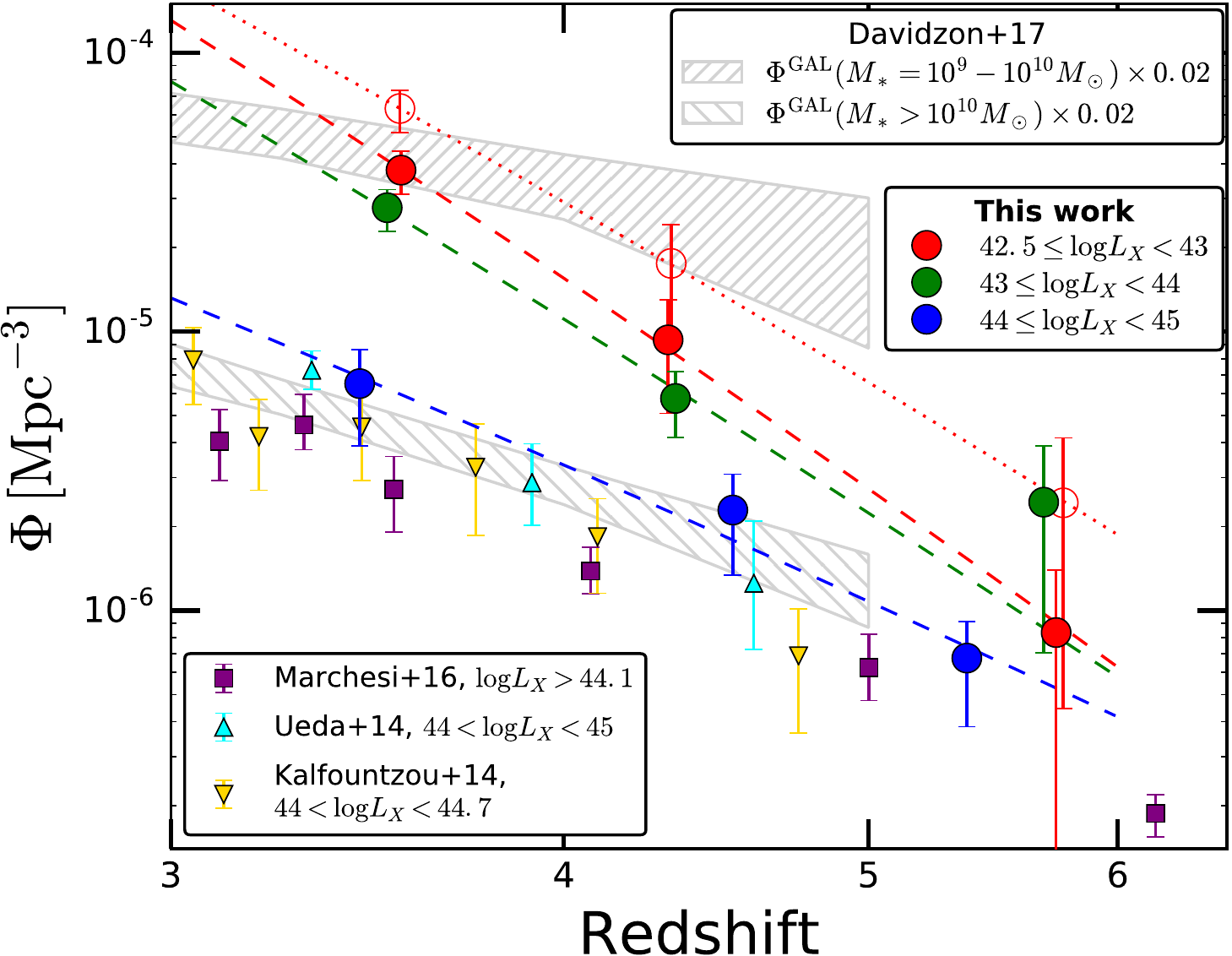}
\caption[]{
{\it Left:\/}
Number-density evolution at $z=0$--5 of AGNs within different \xray\ luminosity 
ranges from the best-fit XLF model of Aird et~al.\ (2015).
The shaded regions indicate model uncertainties (99\% confidence level).
The data points are from Miyaji et~al.\ (2015).
The number-density evolution shows a clear ``downsizing'' pattern, such that 
more luminous AGNs peak at higher redshifts. 
From Aird et~al.\ (2015).
{\it Right:\/}
High-redshift AGN number-density evolution.
The solid large circles are based on the observed AGNs in Vito et~al.\ (2018), 
with different colors indicating the best-fit models. 
The lowest luminosity bin is moderately affected by incompleteness, and the empty
circles are corrected for this effect. 
For higher luminosity bins, the incompleteness effect should be negligible. 
The smaller data points are from different studies as labeled. 
The gray stripes represent galaxy number densities, scaled by a factor of 0.02.
The number density of AGNs at all luminosities decreases sharply toward higher redshifts  
at $z>3$, similarly to massive galaxies. 
From Vito et~al.\ (2018).}
\label{fig:density}
\end{figure}


High-redshift AGNs attract particular attention, because they are closely
related to the seeding and early formation of SMBHs. 
One of the most important lessons from \chandra\ and \xmm\ is that the AGN number density
declines sharply as a function of redshift at $z\gtrsim 3$ (e.g., Barger et~al.\ 2003;
Silverman et~al.\ 2008), which was not shown with the \rosat\ data (see above).
Fig.~8 (right) displays the AGN number density evolution at $z>3$.
Adopting a power-law functional form of $(1+z)^{-\gamma}$, the power-law index describing
the number-density decline is $\gamma \approx 6$--9, depending on the AGN luminosity.
This means there are $\approx 30$--150 times fewer AGNs at $z=6$ than at $z=3$. 
\xray\ stacking analyses indicate that low-luminosity AGNs undetected by the deepest 
\chandra\ surveys contribute negligibly to the total cosmic accretion power
(e.g., Vito et~al.\ 2016).
Given the weak AGN activity at high redshifts, AGNs are unlikely the dominant 
sources driving cosmic hydrogen reionization (e.g., Parsa et~al.\ 2018; Vito et~al.\ 2018; 
Ananna et~al.\ 2020).

While \chandra\ and \xmm\ sample X-ray photons up to $\approx 10$~keV, 
\nustar\ and \swift~BAT can detect photons at higher energies ($\approx3$--79~keV for 
\nustar; 14--195~keV for \swift~BAT).
This allows them to detect and characterize some heavily obscured AGNs, 
which could be very faint in lower-energy bands (e.g., Alexander et~al.\ 2013; 
Civano et~al.\ 2015; Harrison et~al.\ 2016; Lansbury et~al.\ 2017).
Studies based on the \nustar\ and \swift~BAT surveys suggest that the Compton-thick 
($N_{\rm H} \simgt 10^{24}$~cm$^{-2}$) fraction among
all AGNs is $\approx 60\%$, a few times higher than some earlier results
(e.g., Ananna et~al.\ 2019).
Fig.~9 (left) compares the Compton-thick fractions from \nustar\ and \swift~BAT 
vs.\ those from different XLF models.

The ongoing \erosita\ all-sky survey (see \S1.2) will be particularly helpful for 
constraining the XLF at $z\approx 0$ and the luminous end of high-redshift XLF 
(e.g., Wolf et~al.\ 2021).
The \erosita\ team has released a $\approx 140$~deg$^2$ survey field, eFEDS (see 
Fig.~2 left; Brunner et~al.\ 2021).
Fig.~9 (right) compares the $z\approx 6$ XLF constrained by eFEDS with
those extrapolated from lower-$z$ XLF models.
The eFEDS data point appears to be higher than most of the model values, but 
the uncertainty is still too large for a firm conclusion.
The full \erosita\ release in the future will clarify whether the discrepancy
is statistically significant or not.


\begin{figure}[t!]
\includegraphics[width=0.55\linewidth,angle=0]{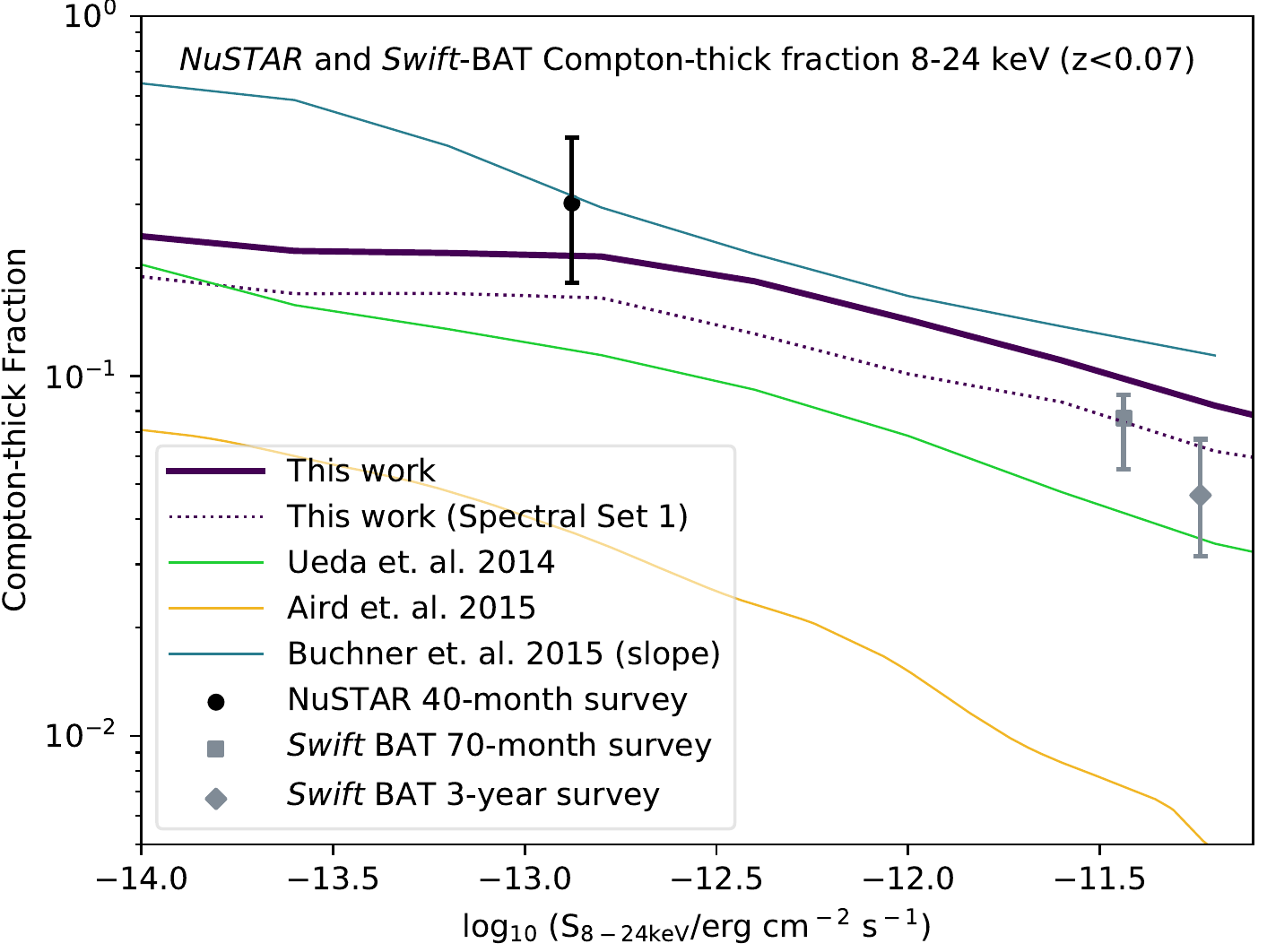}
\includegraphics[width=0.45\linewidth,angle=0]{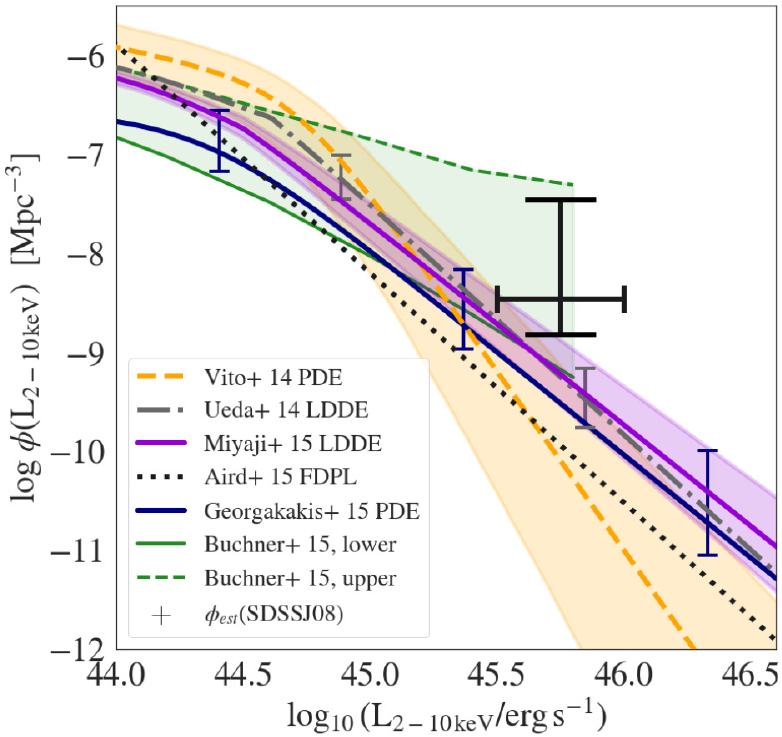}
\caption[]{
{\it Left:\/}
The Compton-thick fraction among all $z<0.07$ AGNs as a function of 8--24~keV flux.
The data points are estimated from \nustar\ or \swift~BAT data.
Different curves are derived from different XLF models as labeled. 
\nustar\ and \swift~BAT data are helpful for constraining the Compton-thick 
population.
From Ananna et~al.\ (2019).
{\it Right:\/}
The \xray\ luminosity function at $z\approx 6$.
The thick black data point is derived from eFEDS. 
The curves represent model-extrapolated XLFs from the literature as labeled,
with uncertainties indicated by the colored shaded areas and error bars.
The eFEDS data point appears to be higher than most of the model values.
The full \erosita\ release will clarify if this possible discrepancy
is statistically significant.
From Wolf et~al.\ (2021).
}
\label{fig:ctk}
\end{figure}


\vspace{0.1in}

\textit{X-ray missed AGNs:}
Although \xray\ observations have several advantages for AGN detection (see \S2.2), 
even deep CXRB surveys may still miss some AGNs powered by actively growing SMBHs.
Multiwavelength identifications of these \xray\ missed AGNs are critically important, 
because complete AGN samples are necessary to draw an unbiased picture of SMBH formation
and evolution.
There are mainly two types of AGNs that might be missed by \xray\ selection: 
one is intrinsically \xray\ weak, and the other is heavily obscured. 
We discuss these two types of objects separately below. 

Some luminous quasars like PHL~1811 have been found to have weak \xray\ 
emission, more than 10~times below the expectations from the $L_{X}-L_{UV}$ relation 
for typical AGNs (e.g., Leighly et~al.\ 2007; Luo et~al.\ 2014).
The weak \xray\ emission appears to be intrinsic, as spectral analyses suggest these 
particular quasars are \xray\ unobscured.
However, such \xray\ weak objects appear rare, consisting of only a few percent of SDSS 
quasars (e.g., Gibson et~al.\ 2008; Pu et~al.\ 2020).
Considering this rarity, our general understanding of AGN demographics appears largely 
unaffected by the intrinsically \xray\ weak population.

Compared to intrinsic \xray\ weakness, heavy obscuration is a more common cause 
of incompleteness of \xray\ surveys.
When obscuration reaches above the Compton-thick level,
even hard \xrays\ are significantly suppressed due to Compton scattering and 
subsequent photoelectric absorption.
Despite AGN-demographics studies to correct for the incompleteness
caused by selection biases against Compton-thick sources, these efforts must 
make strong assumptions about several key factors such as $N_{\rm H}$ distribution above 
$10^{24}$~cm$^{-2}$ and \xray\ spectral shapes (e.g., Ueda et~al.\ 2014; Aird et 
al.\ 2015; Buchner et~al.\ 2015; Ananna et~al.\ 2019).
The uncertainties associated with these assumptions are a major obstacle to achieving
the full picture of AGN demographics, which could be quite different from our 
current understanding.
For example, the currently observed sharp decline of AGN activity at $z\gtrsim 3$ 
(see Fig.~8 right) might be exaggerated by Compton-thick obscuration 
(e.g., Yang et~al.\ 2021b).

To address better the issue of \xray\ incompleteness, it is important to develop
multiwavelength identification methods to effectively select Compton-thick AGNs
(see Brandt \& Alexander 2015; Padovani et~al.\ 2017; Hickox \& Alexander 2018 
for reviews).
One popular method is using broad-band mid-infrared (MIR) photometry to capture AGN dust 
re-emission (e.g., Lacy et~al.\ 2004; Donley et~al.\ 2012).
Another is using optical/IR spectroscopy to detect high-ionization lines that 
are beyond typical star-formation ionization energies (e.g., Steidel et~al.\ 2002; 
Alexander et~al.\ 2008).
However, due to the constraints of survey sensitivity and/or wavelength coverage, 
these methods are generally limited to bright objects. 
Fortunately, \textit{JWST} will soon provide unprecedentedly deep IR imaging and spectroscopy.
It may potentially detect a significant population of Compton-thick AGNs and improve
our understanding of AGN demographics across cosmic history 
(e.g., Satyapal et~al.\ 2021; Yang et~al.\ 2021a,b).

\subsection{AGN Physics}
\label{sec:phy}

\textit{X-ray obscuration}:
In \S2.2, we briefly introduced AGN \xray\ obscuration including its measurement,  
the basic $N_{\rm H}$ distribution, and its relation to optical obscuration.
In this section, we mainly focus on the dependence of \xray\ obscuration upon
redshift, AGN luminosity, and host-galaxy properties.


AGN \xray\ obscuration shows significant cosmic evolution.
The general trend is that the obscuration rises strongly from $z=0$ to $z\approx 2$ 
(e.g., Buchner et~al.\ 2015; Liu et~al.\ 2017).
Fig.~10 (left) displays the obscured AGN fraction as a function of redshift. 
For AGNs with $10^{43.5}<L_X<10^{44.2}$~erg~s$^{-1}$, the obscured fraction
($N_{\rm H} > 10^{22}$~cm$^{-2}$) increases from $\approx 40\%$ at $z=0$ to 
$\gtrsim 80\%$ at $z=2$.
The obscuration-redshift relation is broadly consistent with the fact that the 
obscuring materials (mainly cold gas) are more abundant in the earlier universe.
Interestingly, most studies of the Compton-thick AGN fraction
have not found a significant redshift dependence (e.g., Buchner et~al.\ 2015; 
Ananna et~al.\ 2019; Li et~al.\ 2019; but also see Lanzuisi et~al.\ 2018).
However, the uncertainties are large, especially 
considering detection incompleteness for the Compton-thick population 
(e.g., Hickox \& Alexander 2018).
At high redshifts of $z\gtrsim 2$, $N_{\rm H}$ measurements become challenging 
due to the general faintness of high-redshift objects. 
However, studies have suggested that the obscuration-redshift relation ``saturates'' at 
high redshifts, i.e., the obscured fraction roughly remains constant at $z\gtrsim 2$ 
(e.g., Liu et~al.\ 2017; Vito et~al.\ 2018).

At a given redshift ($z\simlt 3$), AGN \xray\ obscuration becomes weaker toward higher luminosities
(e.g., Merloni et~al.\ 2014; Liu et~al.\ 2017).
For example, at $z\approx 1.5$, the obscured fraction drops from 75\% at 
$L_X \approx 10^{43}$~erg~s$^{-1}$ to 40\% at $L_X \approx 10^{44.5}$~erg~s$^{-1}$.
This luminosity dependence can be understood within the framework of the AGN-unification 
model, which envisages a toroidal obscuring structure 
(e.g., Antonucci 1993; Netzer 2015).
In this model, whether a specific AGN is obscured depends on the viewing angle, 
and thus the obscured fraction is numerically equivalent to the torus covering factor
for an AGN sample.
It has been reasonably proposed that the torus covering factor is negatively related 
to AGN luminosity (or Eddington ratio), because more powerful central engines could 
blow away more obscuring materials via radiation pressure and/or winds 
(a ``receding torus''; e.g., Lawrence 1991; Netzer 2015; Ricci et~al.\ 2017). 
However, observations suggest that the obscuration-luminosity anticorrelation may
not hold for high-redshift ($z\simgt 3$) AGNs, although the uncertainties are
considerable (e.g., Vito et~al. 2014, 2018; Buchner et~al. 2015).

Since AGN activity is closely related to host-galaxy properties (see \S3.3), it is 
natural to wonder if AGN obscuration also depends upon galaxy properties. 
Studies have not found a significant difference in host-galaxy $M_*$ or SFR
for \xray\ obscured vs.\ unobscured AGNs (e.g., 
Merloni et~al.\ 2014; Mountrichas et~al.\ 2021b; but also see Lanzuisi et~al.\ 2017).
Interestingly, studies indicate that optical type~2 AGNs tend to have higher $M_*$
compared to type~1 AGNs, although these two classes have similar SFR 
distributions (e.g., Zou et~al.\ 2019; Mountrichas et~al.\ 2021c).
The different behaviors of \xray\ and optical obscuration suggest that they may have 
different physical origins.
However, robust measurements of host-galaxy properties can be challenging, especially
for luminous AGNs which outshine host starlight, and further investigations are needed
to test AGN obscuration-galaxy connections.

\vspace{0.1in}

\textit{Multiwavelength energy distribution of X-ray, UV/optical, and IR}:
Modern imaging surveys can effectively sample the SEDs for large numbers of extragalactic 
sources.
The investigation of AGN multiwavelength SEDs is an effective way to gain insight
into AGN physics, as they reflect the physical structures of, e.g., the 
accretion disk and its corona.

Perhaps the most fundamental SED feature is the tight connection between UV/optical 
and \xray\ luminosities for type~1 AGNs.
Conventionally, researchers have used a quantity, \aox, the SED slope between
UV and \xray.\footnote{\aox$=0.3838 \log \frac{\lxr}{\luvr}$, where $\luvr$ and 
$\lxr$ are rest-frame 2500~$\AA$ and 2~keV monochromatic luminosities with units of 
erg~s$^{-1}$~Hz$^{-1}$, respectively.}
It has been found that \aox\ is tightly correlated with $\luvr$ (e.g., Steffen et 
al.\ 2006; Just et~al.\ 2007; Lusso et~al.\ 2010; Lusso \& Risaliti 2016).
Fig.~10 (right) displays an example of this relation.
The \aox-$\luvr$ relation has an intrinsic scatter of only $\approx 0.1$ in \aox, 
over more than five orders of magnitude in $\luvr$.
This scatter translates to a factor of $\approx 2$--3 in terms of the
$\lxr$-$\luvr$ relation.
The best-fit \aox-$\luvr$ relation has a negative slope of $\approx -0.14$, indicating 
that more luminous AGNs tend to have larger UV-to-X-ray ratios.
Studies have generally not found a significant redshift evolution of the \aox-$\luvr$ 
relation up to $z \approx 6$--7, indicating this relation is determined by some fundamental 
accretion physics (e.g., Vito et~al.\ 2019; Wang et~al.\ 2021; see Fig.~10 right).
Taking advantage of this apparent redshift independence, some researchers have proposed
to use quasars as ``standard candles'' to constrain cosmological parameters 
(e.g., Lusso \& Risaliti 2017). 

The \aox-$\luvr$ relation discussed above is mainly investigated using type~1 AGN samples, 
where the AGN UV emission is directly observable.  
Under the scheme of AGN unification, this relation should also hold for the central 
engines of the type~2 population, despite their obscuration.
Based on this idea, Yang et~al.\ (2020) developed a SED-fitting code, {\sc x-cigale}.
{\sc x-cigale} builds SED templates from \xray\ to IR by applying the \aox-$\luvr$ 
relation to physical AGN UV-to-IR models, in which the intrinsic $\luvr$ is also 
known for type~2 AGNs. 
Using {\sc x-cigale}, studies have successfully fitted observed type~2 SEDs, 
supporting the AGN-unification scheme (e.g., Yang et~al.\ 2020; Mountrichas et 
al.\ 2021a; Toba et~al.\ 2021). 
The fitted X-ray vs.\ MIR AGN luminosities are also consistent with empirical AGN 
$L_X$-$L_{\rm 6 \mu\rm m}$ relations (see, e.g., Fig.~15 of Toba et~al.\ 2021), 
indicating these relations are physically driven by the \aox-$\luvr$ relation.

Some observational studies have suggested that the observed \aox-$\luvr$ relation 
might actually be driven by a more fundamental connection between \aox\ and Eddington 
ratio (e.g., Lusso et~al.\ 2010; Jin et~al.\ 2012). 
The Eddington-ratio dependence is also favored by some theoretical studies of the AGN 
disk-corona system (e.g., Cao 2009; Kubota \& Done 2018).
However, the dependence on Eddington ratio is still controversial, likely due to the 
strong degeneracy with $\luvr$ and the relatively large uncertainties in SMBH-mass 
measurements (e.g., Shemmer et~al.\ 2008; Liu et~al.\ 2021b).


\begin{figure}[t!]
\includegraphics[width=0.53\linewidth,angle=0]{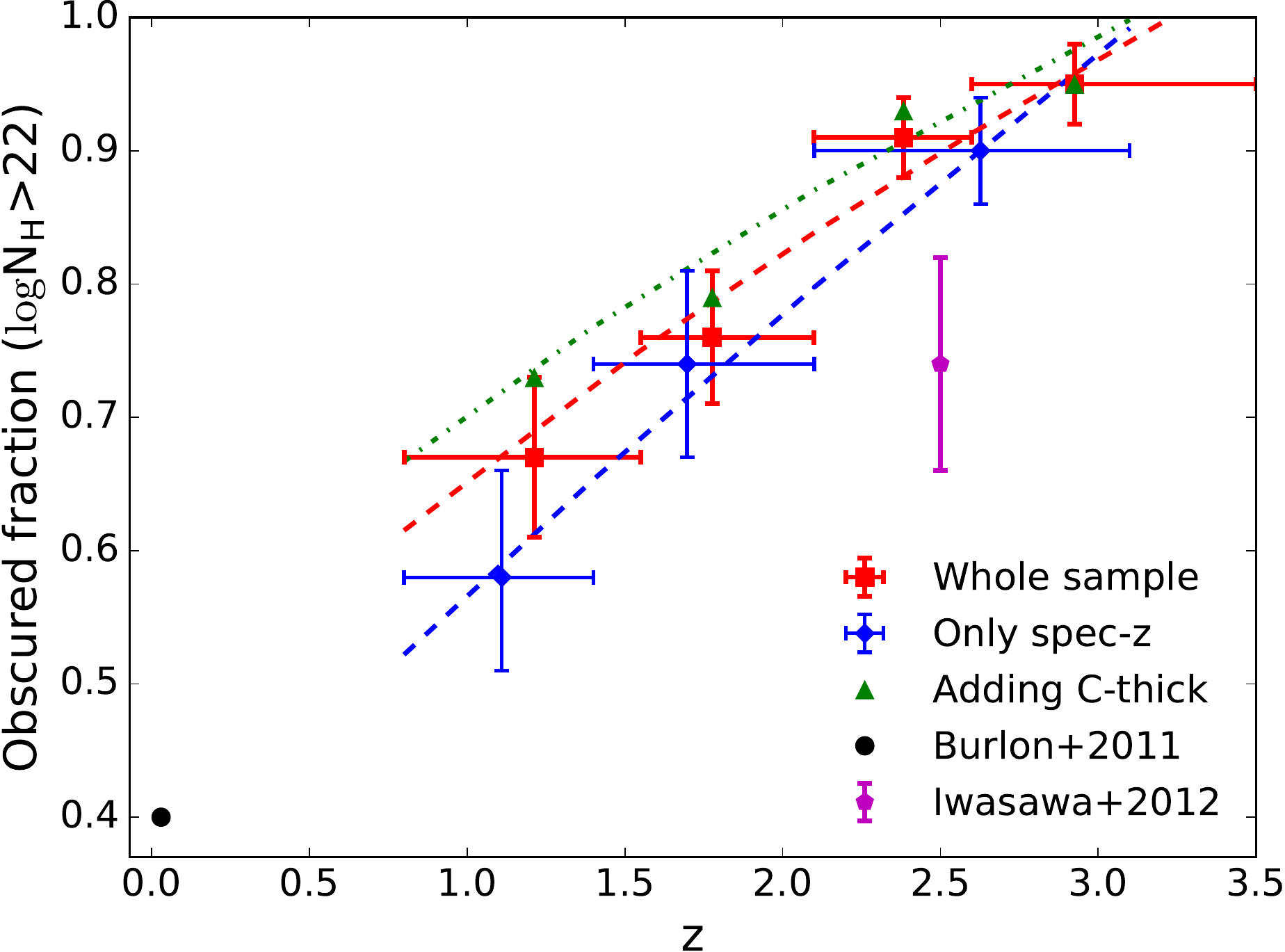}
\includegraphics[width=0.47\linewidth,angle=0]{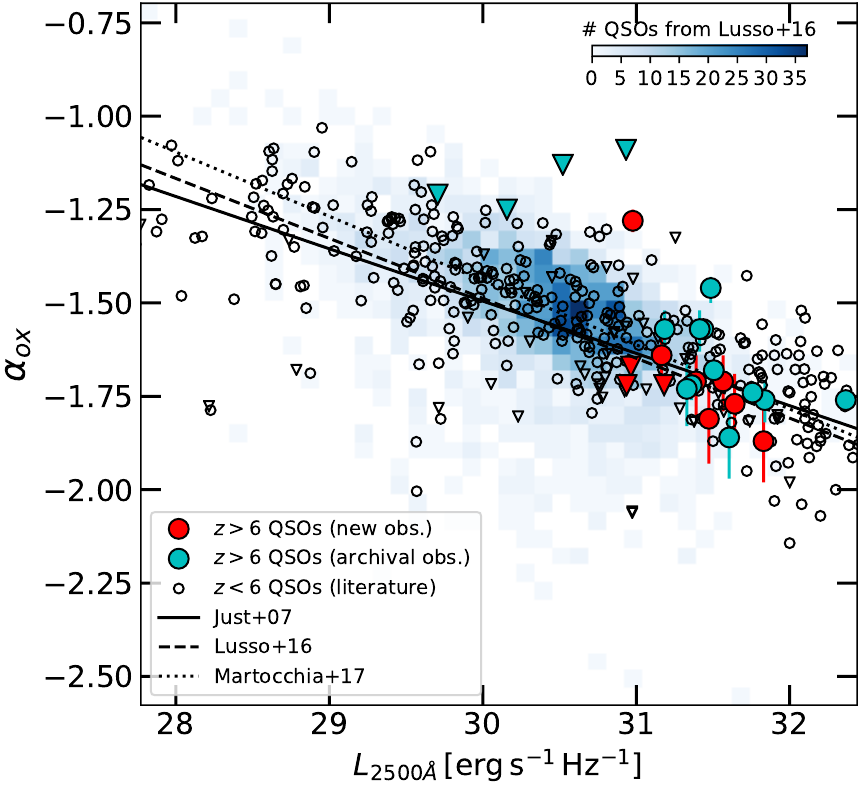}
\caption[]{
{\it Left:\/}
Redshift evolution of \xray\ obscured AGN fraction at $43.5<\log L_X<44.2$ 
(units: erg~s$^{-1}$).
The red data points are from a complete Compton-thin ($N_{\rm H}<10^{24}$~cm$^{-2}$) 
sample, and the red curve is the best fit.
The blue points and curve are from a subsample with spectroscopic redshifts; 
the green points and curve are from the sample plus Compton-thick objects.
The black and purple points are from the literature.
The obscured fraction rises sharply from $z=0$ to $z\approx 2$--3. 
From Liu et~al.\ (2017).
{\it Right:\/}
\aox-$\luvr$ relation.
The larger data points represent $z>6$ quasars, while the smaller data points
and blue shaded regions represent $z<6$ quasars.
The downward-pointing triangles indicate \aox\ upper limits.
The lines are the best-fit relations from the literature as labeled.
The $z>6$ quasar sample is consistent with these \aox-$\luvr$ relations,
which were derived from $z<6$ quasars.
From Vito et~al.\ (2019).
}
\label{fig:obsc}
\end{figure}


\vspace{0.1in}

\textit{X-ray variability}:
Modern \xray\ surveys often obtain multiple exposures on the same sky region to 
reach higher sensitivities.
As a bonus, multiple exposures also enable time-domain analyses on timescales from 
days to years. 
These analyses can effectively reveal some essential physical properties of AGNs.
It has been proposed that a significant fraction of observed AGNs might be powered 
by TDEs instead of normal gas accretion (e.g., Milosavljevi\'c et~al.\ 2006). 
Whether an X-ray AGN is powered by a TDE or gas accretion can be distinguished 
with long-term ($\sim$years) variability analyses.
A TDE-powered AGN likely will decay monotonically over time (except for the initial 
burst), and the change in \xray\ flux can reach orders of magnitude (e.g., Komossa 2015).
In contrast, a normal AGN has stochastic variability with much milder amplitudes.
The long-lived \chandra\ and \xmm\ missions allow the characterization of AGN long-term 
variability.

Fig.~11 (left) compares the fluxes from two observational epochs 
separated by $\approx 15$~years. 
The flux changes between the two epochs are stochastic with amplitudes generally 
less than a factor of $\approx 2$--3.
This result suggests these AGNs are likely powered by normal gas accretion rather
than TDEs. 
Actually, systematic searches for TDEs in \xray\ surveys indicate the fraction of 
TDE-powered AGN is at most \hbox{$\approx 1\%$} (e.g., Luo et~al.\ 2008; Zheng et~al.\ 2017).
The mild stochastic variability also indicates that novel AGN phenomena such as 
``changing-look'' events and emission-state changes are rare, because those phenomena
are often associated with large changes in \xray\ flux.


\begin{figure}[t!]
\includegraphics[width=0.47\linewidth,angle=0]{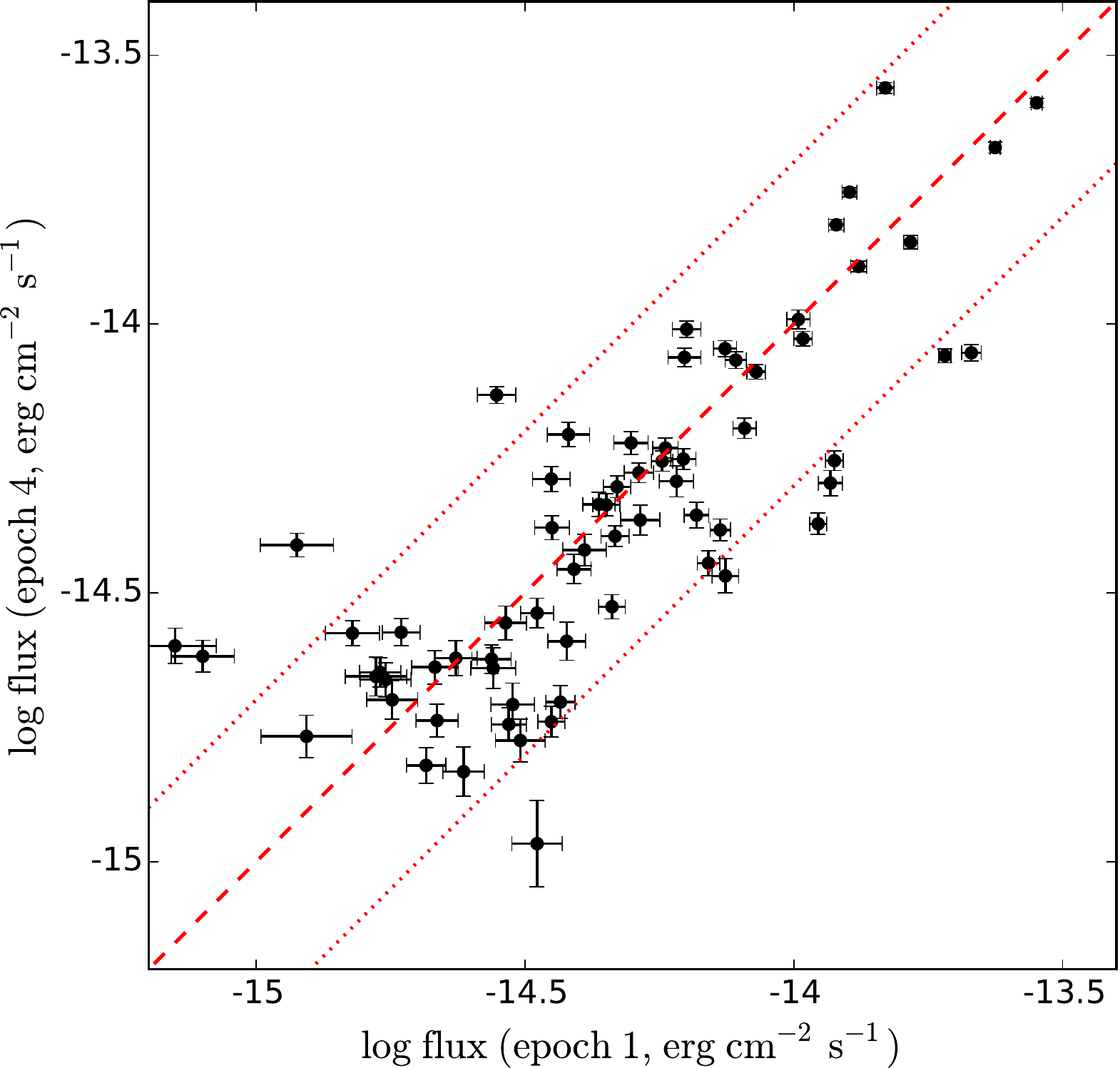}
\includegraphics[width=0.53\linewidth,angle=0]{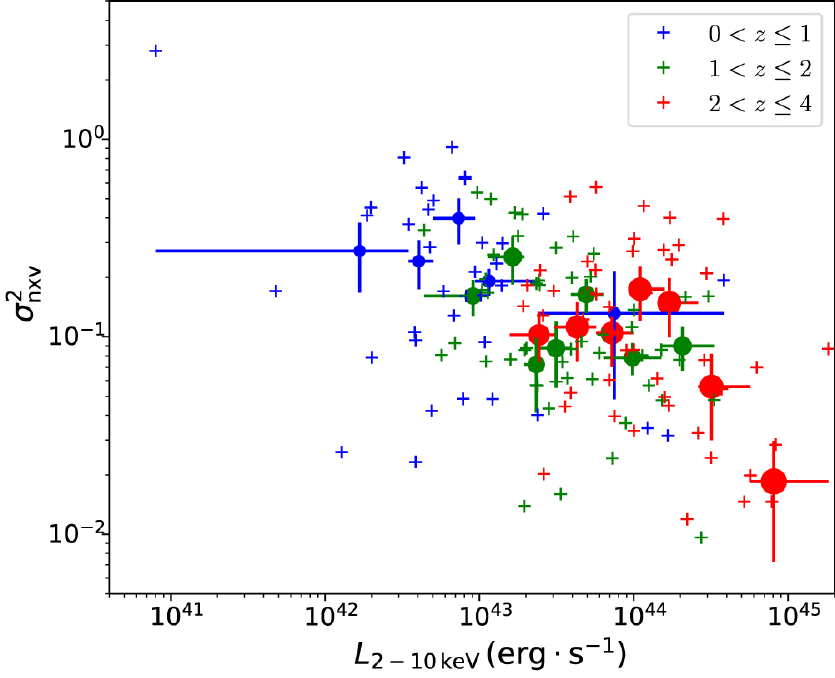}
\caption[]{
{\it Left:\/}
Observed AGN 0.5--7~keV flux comparison for two observational epochs of the 
\chandra\ Deep-Field South separated 
by 15 years in the observed frame.
The AGNs are at redshifts of $\approx0.6$--3.1 (10th--90th percentile).
The red dashed and dotted lines mark no variation and factor-of-two variations,
respectively.
The flux variability appears to be stochastic, and the amplitudes are 
$\lesssim 2$--3. 
Therefore, extreme events (e.g., TDEs and changing-look AGNs) are rare even on 
a 15-year timescale.
From Yang et~al.\ (2016).
{\it Right:\/}
AGN variability strength (normalized excess variance) vs.\ 2--10~keV luminosity.
Different colors indicate different redshift ranges as labeled.
The small crosses represent individual sources.
The large data points indicate the binned results.
AGN variability becomes weaker toward higher luminosities.
From Zheng et~al.\ (2017).
}
\label{fig:var}
\end{figure}


From \xray\ spectral analyses, it has been found that the observed distant AGN \xray\ 
flux variations are largely due to changes of the intrinsic luminosity rather than 
obscuration (e.g., Yang et~al.\ 2016; Falocco et~al.\ 2017). 
This finding means \xray\ variability can be used to probe the central engine
properties.
Deep-field observations have found \xray\ variability amplitudes are anti-correlated 
with luminosities, as is known in the local universe (e.g., Lanzuisi et~al.\ 2014; 
Yang et~al.\ 2016; Zheng et~al.\ 2017).
Fig.~11 (right) displays this anti-correlation.
By fitting the variability-luminosity relation with physical models of power 
spectral density (PSD), researchers can constrain SMBH masses and Eddington ratios 
for different samples and thereby study the cosmic evolution of these fundamental 
properties (e.g., Paolillo et~al.\ 2017; Zheng et~al.\ 2017). 

Finally, timing analyses are a useful way to select AGNs. 
Deep \xray\ surveys can detect faint AGNs that have a similar luminosity level 
($L_X < 10^{42}$~erg~s$^{-1}$) to normal galaxies. 
It is often challenging to select these AGNs with \xray\ or multiwavelength methods 
(see \S2.1) due to their similarity to normal galaxies.
However, if they are found to have significant \xray\ variability, 
then their AGN nature is revealed, because typical normal galaxies have relatively 
constant light curves.
Deep-field studies have successfully used this method to identify low-luminosity AGNs,
some of which are missed by other selection methods (e.g., Young et~al.\ 2012; 
Ding et~al.\ 2018).
Future powerful observatories such as \athena\ and \lynx\ (see \S4.1) will detect many
more faint \xray\ sources than we have today, and the variability selection method 
will be crucial for selecting low-luminosity AGNs among these sources.

\subsection{AGN Ecology}
\label{sec:eco}

From observations of local systems, SMBH masses are tightly correlated with some
host-galaxy properties such as bulge masses and velocity dispersions (e.g., Kormendy 
\& Ho 2013).
These connections are surprising given that SMBH masses are often negligible ($\lesssim$
a few thousandths) compared to the host-galaxy masses.  
Therefore, some physical connections between SMBH growth and host-galaxy properties 
likely exist, referred to here as ``AGN ecology''.
\xray\ surveys and their accompanying multiwavelength data offer an excellent chance 
to study AGN-galaxy connections across cosmic history.
We discuss the topic of AGN-galaxy connections in this section, mainly focusing on 
recent work not covered in Brandt \& Alexander (2015).

\vspace{0.1in}

\textit{Measurements of host-galaxy properties}:
Basic galaxy properties such as $M_*$ and SFR can be estimated by modeling the 
observed broad-band SEDs from the UV to IR.
For optically selected quasars, the AGN component typically outshines the 
host-galaxy component at UV-to-NIR (near-IR) wavelengths, posing a significant 
challenge for SED decomposition.
Fortunately, \xray\ selected AGNs, especially those with low-to-moderate 
luminosities ($L_X \lesssim 10^{44}$~erg~s$^{-1}$) and/or significant nuclear
absorption, often have UV-to-NIR SEDs dominated by starlight 
(see, e.g., Fig.~9 of Luo et~al.\ 2010).
Their $M_*$ and SFR measurements are almost as straightforward as those of normal 
galaxies.
Even for luminous quasars, the \xray\ fluxes can be used to constrain their
AGN component in SED modeling and thereby reduce the uncertainties in the $M_*$ and 
SFR measurements (e.g., Yang et~al.\ 2020; Mountrichas et~al.\ 2021a).

Galaxy morphological measurements are challenging in general.
Since broad-line (BL) AGNs could have a significant contribution to the optical 
imaging, their host-galaxy morphological measurements might be unreliable and/or
heavily biased.
Therefore, morphological studies often focus on non-BL AGN hosts (e.g., Yang et 
al.\ 2019b; Ni et~al.\ 2021b).
Even for non-BL AGNs' host galaxies, morphological measurements are technically 
challenging for ground-based telescopes, as a $1''$ seeing translates to an
$\approx 8$~kpc physical scale at $z=1$, comparable to the typical sizes of $L^*$ 
galaxies.
For this reason, the morphological investigations of \xray\ AGN hosts often rely 
on \hst\ imaging.    

The reddest band ($H_{160}$) from \hst\ allows morphological measurements up to 
$z\approx 3$, above which the covered rest-frame wavelengths shift to the UV and 
the ``morphological k-correction'' becomes strong (e.g., Taylor-Mager et 
al.\ 2007). 
\jwst\ will be able to probe to higher redshifts thanks to its high-resolution 
imaging ability at longer wavelengths ($\approx 0.6$--$5\ \mu$m for NIRCam).
There are various methods of making morphological measurements, e.g., S\'ersic-profile
fitting, concentration-asymmetry-clumpiness (CAS), and visual classifications (see 
Conselice 2014 for a review).
Visual classifications from expert classifiers are often very time consuming.
Fortunately, the advances of machine-learning techniques in recent years allow
effective visual-like morphological classifications for large numbers of distant galaxies 
(e.g., Huertas-Company et~al.\ 2015). 

\vspace{0.1in}

\textit{Host-galaxy stellar mass and star-formation rate:}
As we have discussed in \S3.1, both SMBH growth and galaxy stellar growth 
have the anti-hierarchical behavior of downsizing. 
Also, the cosmic SMBH-accretion density and star-formation density\footnote{
The total SMBH accretion rate and star-formation rate per comoving volume,
both in units of $M_\odot$~yr$^{-1}$~Mpc$^{-3}$.} both peak at $z\approx 2$
(see, e.g., Fig.~20 of Aird et~al.\ 2015).
These similarities suggest a connection between AGN activity and star formation (SF).

One major challenge in studying the AGN-SF connection is ``AGN flickering''.
From hydrodynamical simulations, AGNs are likely associated with strong short-term 
($\lesssim 10$~Myr) variability (e.g., Yuan et~al.\ 2018). 
This temporal behavior acts as ``noise'' in the analyses of SMBH cosmic evolution,
for which only long-term \hbox{($\gtrsim 100$~Myr)} AGN strength is relevant. 
To overcome this variability issue, Chen et~al.\ (2013) and Hickox et~al.\ (2014) 
proposed to use sample-averaged SMBH accretion rate (BHAR, estimated from the 
combination of direct \xray\ detection and stacking) to approximate long-term average BHAR. 
They found that sample-averaged BHAR is linearly correlated with galaxy SFR, 
and interpreted this result as a tight universal connection between SMBH and stellar
growth over galaxy evolution timescales ($\gtrsim 100$~Myr).
In this scenario, long-term BHAR can be simply inferred from host-galaxy SFR alone.

\begin{figure}[t!]
\includegraphics[width=0.33\linewidth,angle=0]{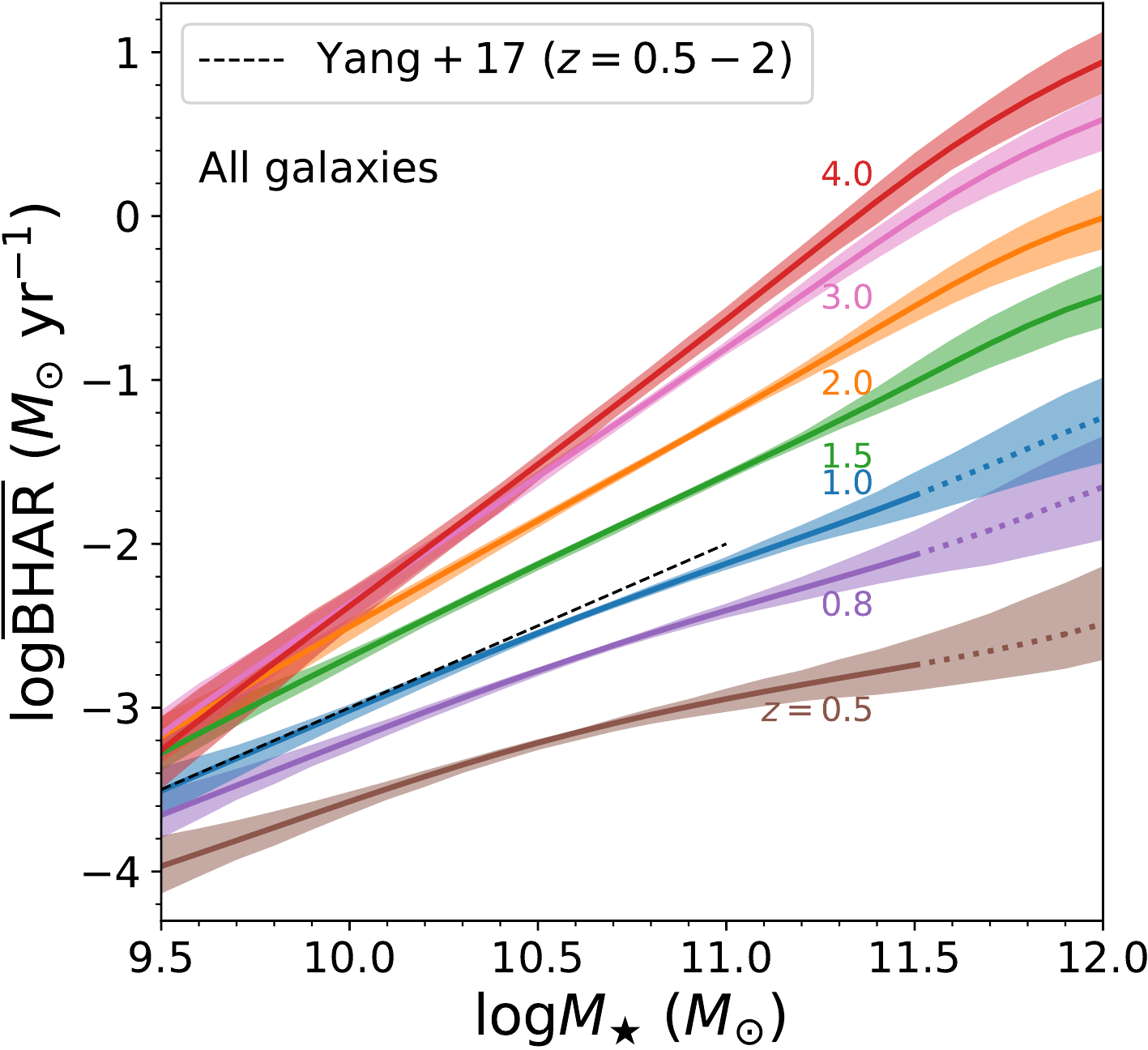}
\includegraphics[width=0.67\linewidth,angle=0]{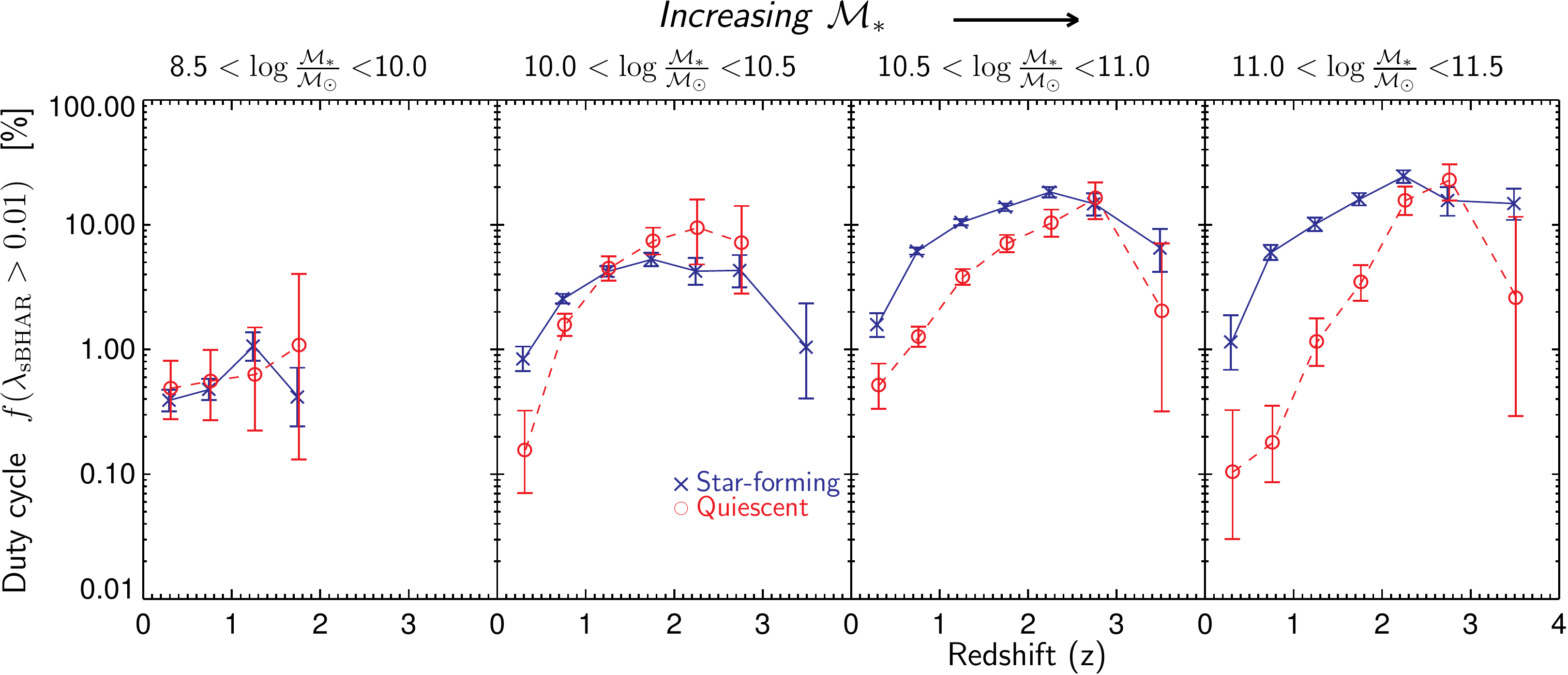}
\caption[]{
{\it Left:\/}
Average BHAR-$M_*$ relation (AGN main sequence) for all galaxies at different redshifts 
(as labeled).
The dashed line represents the best-fit relation based on a $z=0.5$--2 sample from 
the literature.
At any fixed $M_*$, BHAR rises toward high redshifts, but the redshift dependence 
is stronger in more massive galaxies.
From Yang et~al.\ (2018a).
{\it Right:\/}
AGN fraction (duty cycle) as a function of redshift for different $M_*$ bins.
$M_*$ increases from the leftmost to the rightmost panels.
The blue and red data points represent star-forming and quiescent galaxies, 
respectively.
The AGN-fraction difference between star-forming vs.\ quiescent samples depends 
on redshift and $M_*$.
From Aird et~al.\ (2018).}
\label{fig:bhar}
\end{figure}

On the other hand, deep-field studies have also found that the fraction of AGN-hosting 
galaxies rises steeply as a function of stellar mass (e.g., Xue et~al.\ 2010; 
Aird et~al.\ 2012; Bongiorno et~al.\ 2012).
This mass dependence appears to challenge the AGN-SF connection above.  
This apparent paradox could result from the SF main sequence, i.e., SFR and $M_*$ are 
positively correlated among star-forming galaxies at a given redshift.
For example, AGN activity might be primarily related to SFR, while the apparent 
$M_*$ dependence might be a bias due to the SFR-$M_*$ main sequence, or vice 
versa.
To test this idea, Yang et~al.\ (2017) performed partial-correlation analyses,
which control SFR ($M_*$) and investigate the dependence on $M_*$ (SFR).
Their result, somewhat surprisingly, indicates that sample-averaged BHAR is mainly 
correlated with $M_*$, and the apparent BHAR-SFR connection is a secondary relation 
caused by the SF main sequence.
Similar conclusions have also been reached by subsequent studies (e.g., Fornasini 
et~al.\ 2018; Ni et~al.\ 2019).
Observational works have also investigated the effects of host-galaxy cosmic
environments on BHAR when controlling for $M_*$ (e.g, Powell et~al.\ 2018; 
Yang et~al.\ 2018b; Allevato et~al.\ 2019; Noordeh et~al.\ 2020).
The environmental dependence is generally weak, except for the most massive 
clusters.

Motivated by the importance of $M_*$, studies have attempted to quantify the BHAR 
dependence on $M_*$ at different redshifts (e.g., Georgakakis et~al.\ 2017; 
Aird et~al.\ 2018; Yang et~al.\ 2018a).\footnote{The 
BHAR vs.\ $M_*$ relation is sometimes called the ``AGN main sequence'', in analogy with 
the well-known SF main sequence (e.g., Mullaney et~al.\ 2012).}
Such works often have to collect large samples of AGNs, combining both deep and 
wide \xray\ surveys to cover AGNs from low-to-high luminosities.
Fig.~12 (left) shows the resulting BHAR-$M_*$ relation. 
At any fixed $M_*$, BHAR rises toward high redshifts, likely reflecting the fact that 
AGN-accretion fuel (cold gas) is more abundant in the earlier universe. 
However, the redshift dependence is stronger in more massive galaxies, indicating
that massive galaxies assemble their SMBH masses earlier in cosmic history, 
compared to low-mass galaxies.
This phenomenon is broadly consistent with AGN downsizing (see \S3.1), considering 
that more massive galaxies generally have more massive SMBHs.

We caution that AGN activity could still have some ``residual'' dependence on SFR, 
although $M_*$ is the main factor.
Studies have also investigated the AGN dependence on host-galaxy positions with respect to 
the SF main sequence at a given $M_*$ (e.g., Aird et~al.\ 2018, 2019; Masoura et~al.\ 2018; 
Yang et~al.\ 2018a; Bernhard et~al.\ 2019; Florez et~al.\ 2020; Mountrichas et~al.\ 2021b).
Fig.~12 (right) displays this dependence, where galaxies are divided 
into star-forming and quiescent populations.
The resulting SF dependence is quite complicated, apparently affected by both redshift 
and $M_*$.
More detailed studies found evidence suggesting the AGN activity might not be a simple 
monotonic function of host-galaxy star formation.
For example, AGN activity drops from the starburst to the main sequence (MS) populations, 
but it surprisingly appears to rise from the MS to the sub-MS populations (see, e.g., 
Fig.~11 of Aird et~al.\ 2019).
This complexity suggests there might be some ``hidden parameters'' beyond SFR 
and $M_*$ affecting SMBH growth.
These parameters could be related to galaxy morphology as we will discuss below.

\vspace{0.1in}

\textit{Host-galaxy morphology:}
In local systems, SMBH masses are closely related to host-galaxy morphology. 
For example, they are tightly correlated with bulge masses, but are almost 
completely unrelated to galactic-disk masses (see Kormendy \& Ho 2013 for a review).
Therefore, it is likely that AGN activity in the distant universe also depends 
on host-galaxy morphology.

It has been long known that optically and \xray\ selected AGNs can be found in 
galaxies with a variety of morphological types, e.g., bulge-dominated, irregular, 
and disk-dominated, since the first studies on this topic
(e.g., Bahcall et~al.\ 1997; Grogin et~al.\ 2005).
However, early works did not reach a consensus on whether AGN activity
is physically related to host-galaxy morphologies, especially after controlling 
for the effects of stellar mass and redshift as discussed above (e.g., Pierce et 
al.\ 2007; Kocevski et~al.\ 2012).
One major reason for the disagreement among these works are the technical challenges  
in morphological measurements (see above), which significantly limited the 
sample sizes ($\lesssim 100$ \xray\ AGNs).

The situation has been gradually improved with the advance of the \hst\ 
CANDELS survey (Grogin et~al.\ 2011; Koekemoer et~al.\ 2011), which has
provided deep $H_{160}$ imaging over five separate fields (total area 
$\approx 1000$~arcmin$^2$).
Based on the CANDELS data, robust morphological measurements and classifications 
have been performed for $\approx 50,000$ galaxies up to $z\approx 3$ 
(e.g., Huertas-Company et~al.\ 2015).
There are $\approx 1,000$ \xray\ detections among these galaxies, thanks to the 
deep \chandra\ and \xmm\ exposures for the CANDELS fields.

Taking advantage of these large AGN and galaxy samples, Yang et~al.\ (2019b) 
revealed a tight correlation between sample-averaged BHAR and SFR for 
\textit{bulge-dominated} systems, and this BHAR-SFR correlation is largely 
unaffected by $M_*$ (see Ni et~al.\ 2019 for a similar conclusion). 
Fig.~13 (left) shows this correlation.
The best-fit relation has $\rm BHAR/SFR \approx 1/300$, similar to the SMBH/bulge 
mass ratio observed in the local universe (e.g., Kormendy \& Ho 2013).
This similarity indicates that the BHAR-bulge SFR relation could be the origin of 
the SMBH-bulge mass relation in local systems.
Yang et~al.\ (2019b) also found that BHAR is not related to SFR for the 
non-bulge-dominated sample, highlighting the key role of morphology.
The bulge-dominated systems only consist of $\approx 1/4$ of the massive
($\gtrsim 10^{10}\ M_\odot$) galaxy population.
This explains why the strong BHAR-bulge SFR relation was missed by the 
previous works that did not have morphological information (e.g., Yang et~al.\ 
2017; Fornasini et~al.\ 2018).
The BHAR-bulge SFR relation might also hold for the bulge component in 
bulge-plus-disk systems. 
This can be tested if bulge SFR can be separated from disk SFR by future 
high-resolution mid/far-IR imaging by, e.g., \jwst\ and \textit{Origins}.  


\begin{figure}[t!]
\includegraphics[width=0.4\linewidth,angle=0]{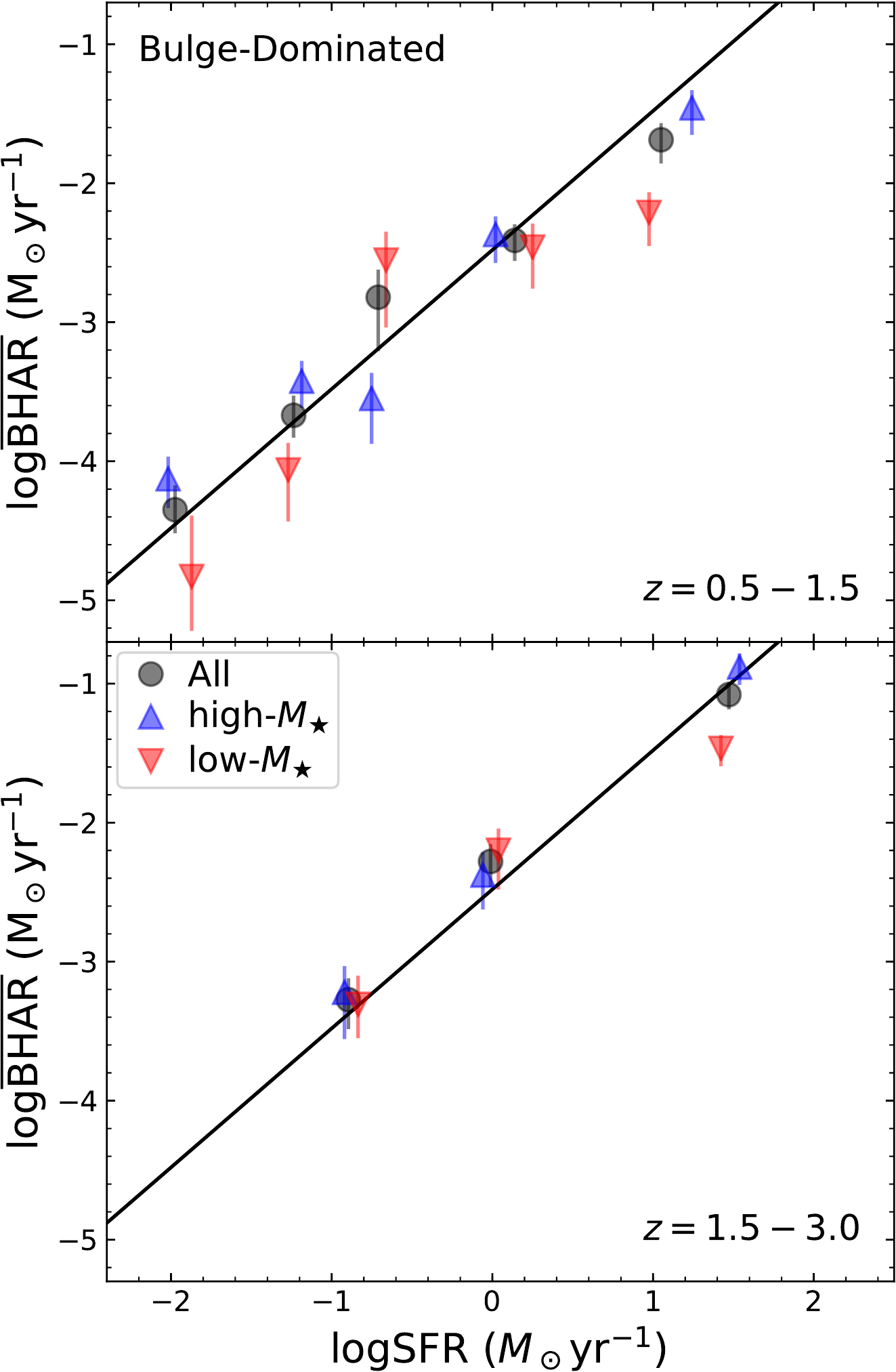}
\includegraphics[width=0.6\linewidth,angle=0]{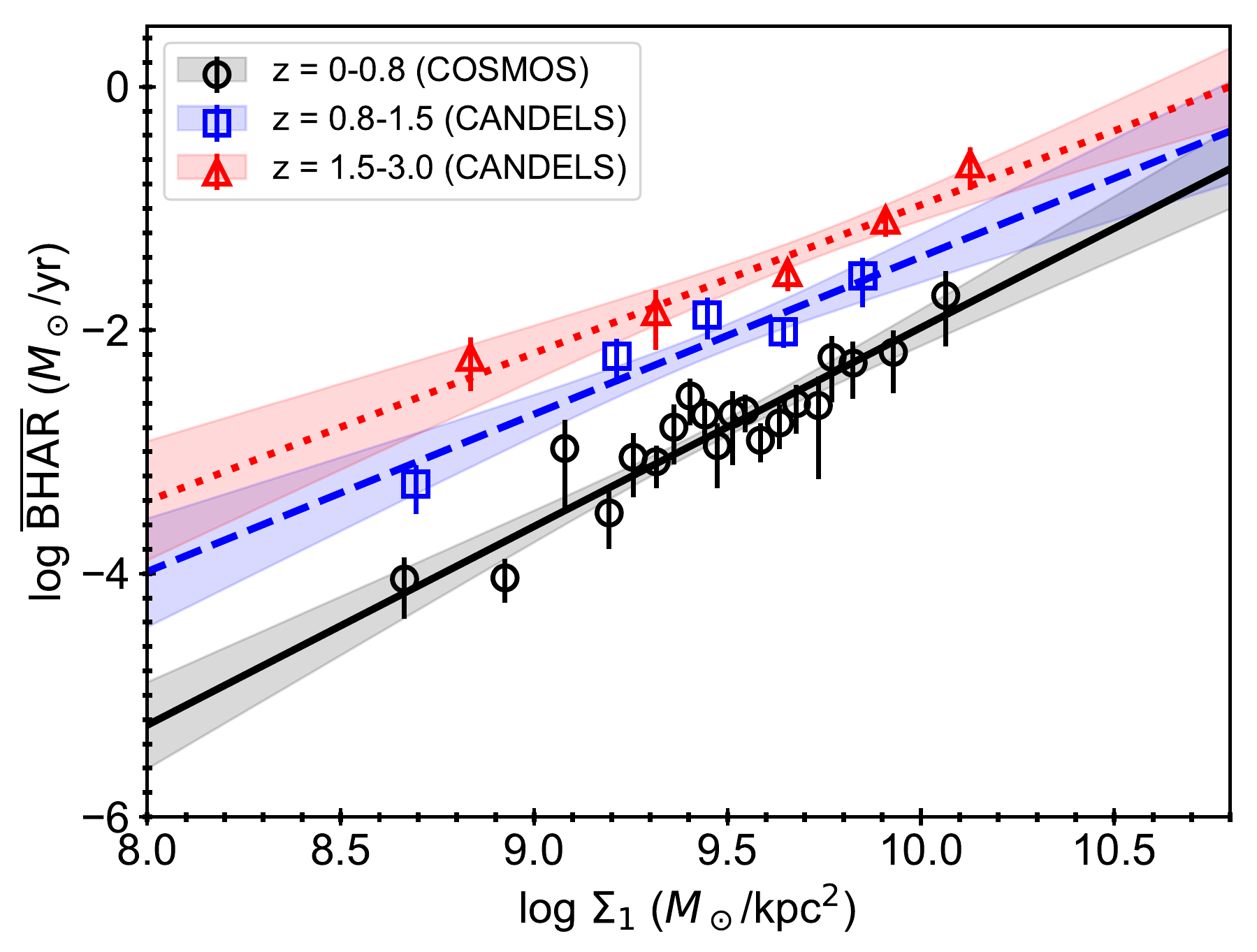}
\caption[]{
{\it Left:\/}
The average BHAR-SFR relation for \textit{bulge-dominated} galaxies.
The top and bottom panels are for different redshift bins.
Each black data point is estimated based on all galaxies in the corresponding SFR 
sample.
To assess the effects of stellar mass, each SFR sample is divided into high-$M_*$ 
(blue) and low-$M_*$ (red) subsamples
The BHAR does not appear to be significantly affected by $M_*$ in general.
The black lines represent the best-fit relation (same in both panels).
The strong BHAR-bulge SFR relation suggests that SMBHs coevolve with bulges,
consistent with observations of local systems.
From Yang et~al.\ (2019b).
{\it Right:\/}
The average BHAR-$\sigone$ relation for \textit{star-forming} galaxies.
Different colors indicate different redshift bins.
The lines and shaded areas are the best-fit models and errors.
In all three redshift bins, BHAR rises toward high $\sigone$.
From Ni et~al.\ (2021b).}
\label{fig:morph}
\end{figure}


Besides morphological classifications, another common way to describe galaxy shapes
utilizes continuous quantities. 
Among these morphological quantities, $\sigone$ (the projected central surface-mass 
density within 1 kpc) has become increasingly popular among researchers, as observations 
have found $\sigone$ is closely related to galaxy evolution (e.g., Barro et~al.\ 2017).
For AGN activity, studies have found AGN fraction is elevated among high-$\sigone$ 
compact star-forming galaxies (e.g., Kocevski et~al.\ 2017; Ni et~al.\ 2019).
This result is qualitatively consistent with the BHAR-bulge SFR relation (Yang et 
al.\ 2019b), because bulge-dominated galaxies are often compact.
Ni et~al.\ (2021b) further quantified the BHAR dependence on $\sigone$ among 
\textit{star-forming} galaxies, as displayed in Fig.~13 (right).
BHAR increases monotonically toward high $\sigone$ for all of the three redshift 
bins investigated.
The conclusion suggests a link between AGN activity vs.\ host-galaxy gas content 
and/or gravitational potential well on the central $\approx\ $~kpc scale, which are 
closely related to the quantity $\sigone$.

Despite the recent significant progress on AGN links to host-galaxy properties 
(especially morphology), we still lack a comprehensive and quantitative description 
of how AGN activity depends on all the key properties including (but not limited to) 
$M_*$, SFR, morphology, and redshift.  
Much larger AGN/galaxy samples are needed to achieve this goal.
The future is bright for this fast-developing field, as many advanced multiwavelength 
survey telescopes will begin operation.
In addition to future \xray\ telescopes (see \S4.1), Rubin will provide deep half-sky 
$ugrizy$ images for photometric redshifts, $M_*$, and SFR measurements.
\jwst\ will open the window for high-redshift galaxies and Compton-thick AGNs (see \S3.1).
\textit{Roman} and \xuntian\ will allow large-scale morphological measurements 
thanks to their efficient high-resolution imaging capability.


\section{\textit{Some Future Prospects and Other Relevant Reviews}}

\subsection{Some Future Prospects for CXRB Surveys}

Many fundamental science questions remain that can be best answered
with ongoing and new CXRB surveys, including the following:

\begin{enumerate}

\item
How do highly obscured SMBHs grow and provide feedback through the \hbox{$z\approx 1$--4}
galaxy formation era?

\item
How do SMBHs grow in the first galaxies at \hbox{$z\approx 4$--10}, and what are the
``seeds'' for these SMBHs? 

\item
What host-galaxy properties link most strongly to SMBH growth, and what does this
reveal about SMBH-galaxy co-evolution? 

\item
How does large-scale cosmic environment, ranging from voids to superclusters, relate to SMBH growth? 

\item
What galaxy physical properties drive the observed cosmic evolution of their
X-ray binary populations, and how do these populations evolve at \hbox{$z\approx 3$--10}? 

\item
What will a $\approx 10^5$-object \xray\ cluster/group sample reveal about the growth of the most
massive cosmic structures, the cosmological model, and dark energy? 

\item
How did the ICM form and SMBHs grow in $z\simgt 2$ protoclusters? 

\item
What is the nature of faint \xray\ transients in the distant universe? 

\end{enumerate}

The CXRB-surveys community is blessed to have multiple \xray\ missions capable of
addressing these questions still operating, including \chandra, \xmm, \integral,
\swift, \nustar, and \srgs. These missions alone should ensure another great
decade of CXRB survey discoveries, particularly if they undertake aggressive projects
capable of exploring new ``discovery space''. Moreover, the vast archives of
\xray\ observations from these missions (e.g., note the locations of the
serendipitous surveys in Fig.~2) will be made immensely more valuable
as the extraordinary new data from complementary imaging and spectroscopic
surveys become available, such as those from
Rubin, \xuntian, SDSS, AAT, DESI, 4MOST, MOONS, PFS, Extremely Large Telescopes,
\jwst, \euclid, \roman, 
ALMA, LMT TolTEC, 
the VLA, ASKAP, MeerKAT, and the SKA.  
These will allow thousands of archival sensitive pointed \xray\ observations
and the all-sky surveys to be systematically mined to address science questions
comprehensively with titanic \xray\ source samples (e.g., Brandt \& Vito 2017). 


\begin{figure}[t!]
%
\includegraphics[height=1.8in,width=2.3in,angle=0]{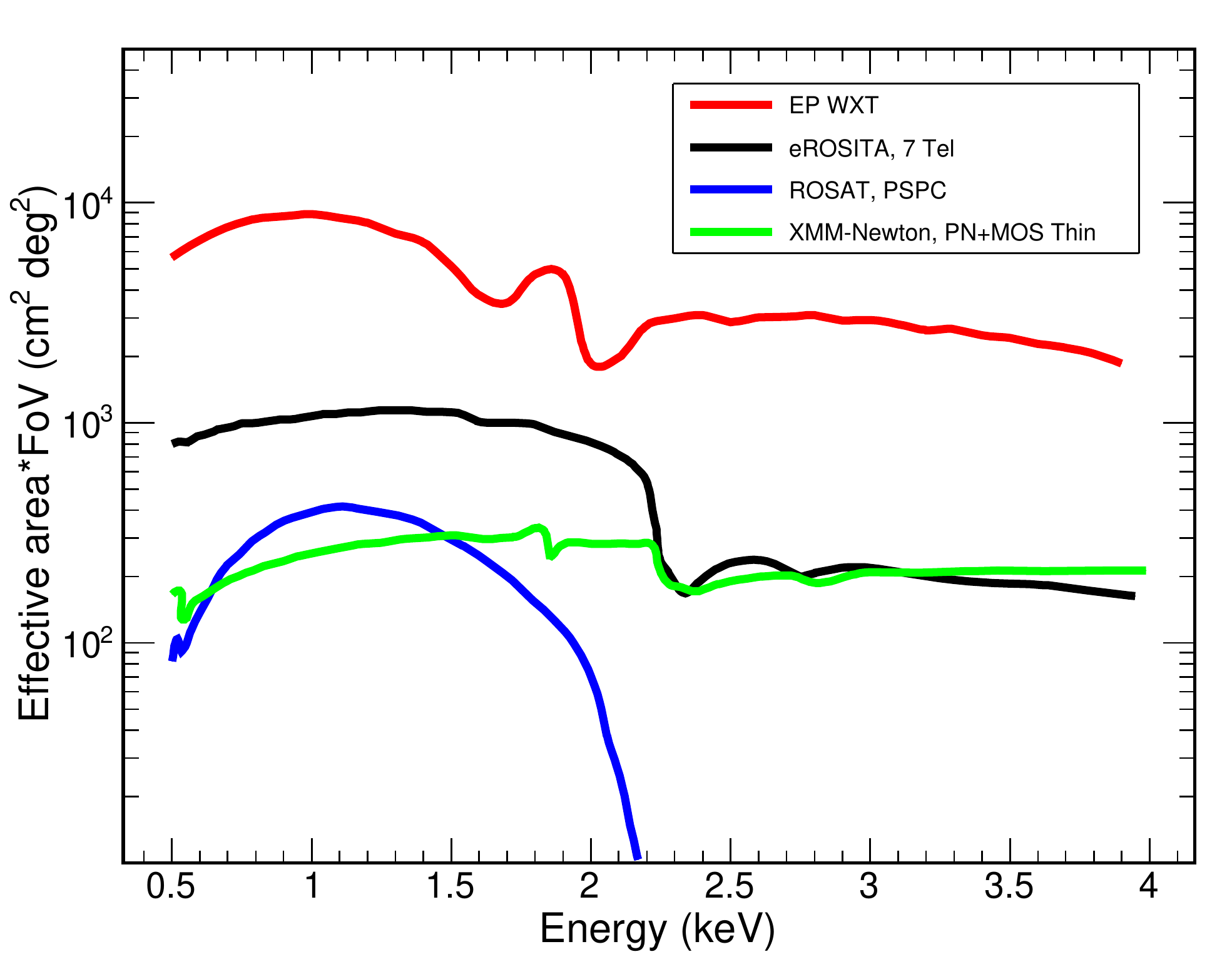}
\includegraphics[height=1.8in,width=2.3in,angle=0]{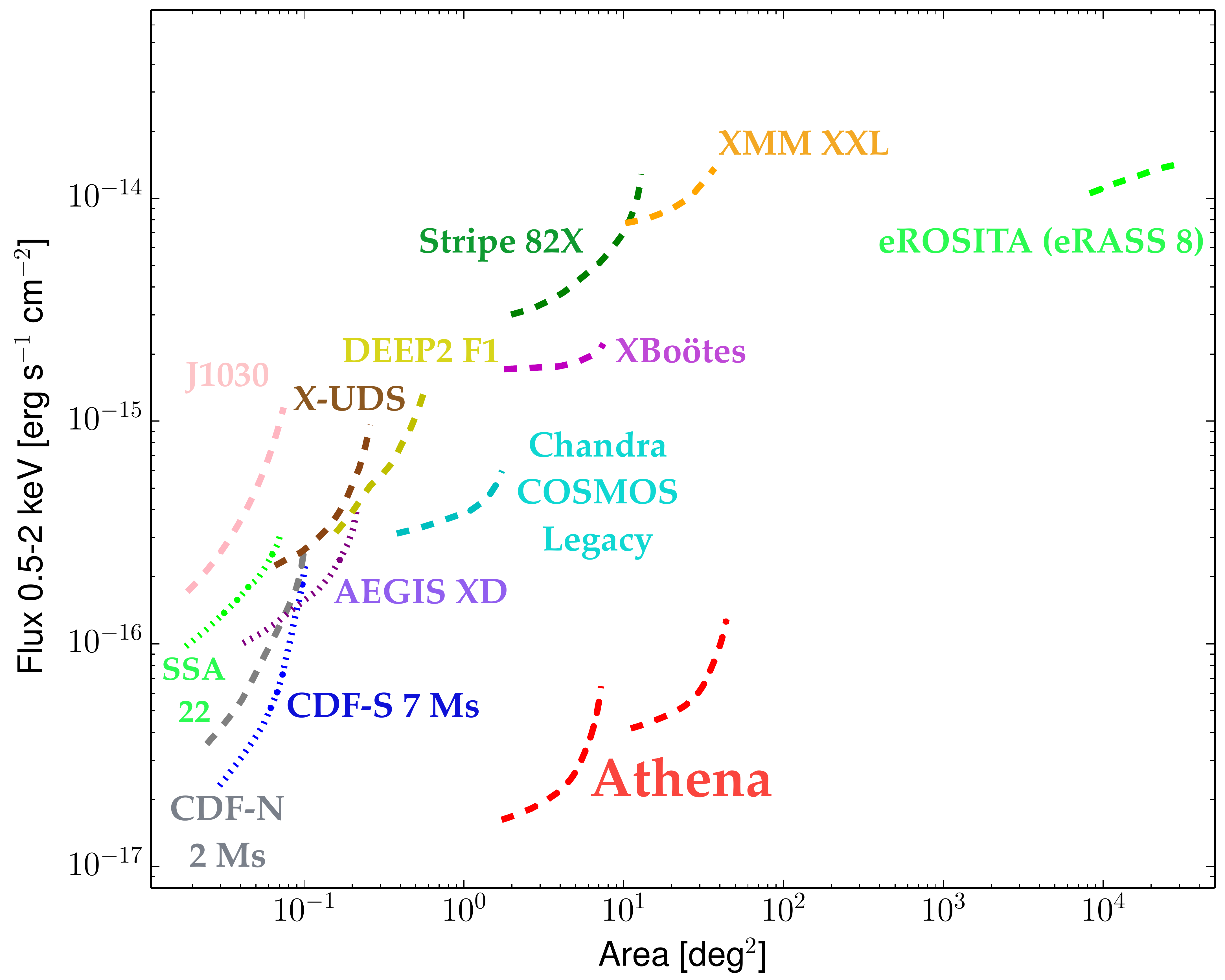}
\caption[]{
{\it Left:\/}
Grasp (the product of effective area and field of view) vs.\ energy for the
\ep\ Wide-Field \xray\ Telescope (EP WXT), compared to the grasps of some current
and previous missions with focusing \xray\ telescopes. The large grasp of \ep\
should make it a powerful wide-field transient survey facility. 
Updated from Zhao et~al.\ (2014). 
{\it Right:\/}
\hbox{0.5--2~keV} flux limit vs.\ area for planned \athena\ CXRB surveys compared
to a selection of previous CXRB surveys (a half-energy width of the \athena\
point spread function of $5^{\prime\prime}$ is assumed). Note the large areas that
\athena\ aims to cover to excellent sensitivity. 
Adapted from Marchesi et~al.\ (2020).}
\end{figure}


\ep, planned for launch in 2023, should greatly advance understanding
of transient sources within the \hbox{0.5--4~keV} CXRB. It will utilize
Lobster-eye optics (covering $\approx 3600$~deg$^2$) to perform a systematic large-grasp
survey of transients at an unprecedented combination of sensitivity and cadence
(e.g., see Fig.~14 left). Discovered transients will be investigated with an on-board
follow-up \xray\ telescope.

\athena, selected for launch in $\approx 2034$, will be a potent mission
for \hbox{$\approx 0.3$--10~keV} CXRB surveys if it achieves its planned
imaging point spread function; e.g., covering several deg$^2$ to
\hbox{0.5--2~keV} fluxes of $\approx 2\times 10^{-17}$ erg~cm$^{-2}$~s$^{-1}$ (see Fig.~14 right)
and providing sufficient photon statistics for quality source characterization. 
\athena\ surveys will hopefully provide a complete census of SMBH growth, including
highly obscured systems, out to \hbox{$z\approx 8$--10}. 

\lynx, a candidate large NASA mission,
aims to observe the seeds of the first SMBHs directly as well as conduct much
superb broader survey science. With \chandra-like angular resolution
and a collecting area \hbox{$\approx 30$--50} times that of \chandra,
\lynx\ could detect $\approx 30,000$~M$_\odot$ black holes at $z\approx 10$
(with \hbox{0.5--2~keV} fluxes of $\approx~5\times 10^{-19}$~erg~cm$^{-2}$~s$^{-1}$;
compare with Fig.~2 left). 
Seeding models (e.g., Population III remnants vs.\ direct-collapse black holes)
would be discriminated using the observed \xray\ luminosity function and host
properties. 

Additional impressive surveyors of the CXRB, including \starx, \axis, and \hbox{\hexp}, 
are under active scientific development and will hopefully soon obtain construction
funding. 

\subsection{Other Relevant Reviews}

Owing to page and reference limits, this overview has been highly concise.
Other useful relevant reviews published over the past decade include
Treister \& Urry (2011), 
Brandt \& Alexander (2015), 
Netzer (2015), 
Xue (2016),
Padovani et~al.\ (2017),
Hickox \& Alexander (2018), 
Allen \& Mantz (2020),
Siemiginowska \& Civano (2020), and
Krivonos et~al.\ (2021). 



\section{\textit{Acknowledgments}}

We thank
DM~Alexander, 
CT~Chen,
L~Klindt, 
BD~Lehmer, 
B~Luo,
Q~Ni,
CJ~Papovich, and
F~Vito
for helpful discussions. 
WNB acknowledges support from
the V.M. Willaman Endowment, 
NSF grant AST-2106990, 
NASA grant 80NSSC19K0961, and
Penn State ACIS Instrument Team Contract SV4-74018
(issued by the Chandra \xray\ Center, which is operated by the
Smithsonian Astrophysical Observatory for and on behalf of
NASA under contract NAS8-03060).
The Chandra ACIS Team Guaranteed Time Observations (GTO) utilized
were selected by the ACIS Instrument Principal Investigator, Gordon P. Garmire,
currently of the Huntingdon Institute for \xray\ Astronomy, LLC, which is under
contract to the Smithsonian Astrophysical Observatory via Contract SV2-82024.
GY acknowledges support from the NASA/ESA/CSA James Webb Space Telescope
through the Space Telescope Science Institute, which is operated by the
Association of Universities for Research in Astronomy, Incorporated,
under NASA contract NAS5-03127. Support for program number JWST-ERS-01345
was provided through a grant from the STScI under NASA contract NAS5-03127.
GY acknowledges support from the George P.\ and Cynthia Woods Mitchell
Institute for Fundamental Physics and Astronomy at Texas A\&M University.




%
%

\end{document}